\documentclass[twoside,11pt]{article}
\usepackage[dvipsnames]{xcolor}

\usepackage{jmlr2e}
\usepackage{lscape}
\usepackage{caption}
\usepackage{array}
\usepackage{lmodern}
\usepackage{multirow}
\usepackage{comment}
\usepackage{enumitem}
\usepackage{url}
\usepackage{xcolor}
\usepackage[framemethod=TikZ]{mdframed}
\usepackage{tfrupee}
\usepackage{amssymb}
\usepackage{amsmath}

\mdfdefinestyle{outbox}{
    linecolor = black,
    linewidth = 1,
    backgroundcolor=white,
    roundcorner = 10pt,
    usetwoside=false,
}

\mdfdefinestyle{inbox}{
    linecolor = black,
    linewidth = 1,
    backgroundcolor=gray!10,
    roundcorner = 10pt,
    usetwoside=false,
    leftmargin = 3.6cm,
}
\begin{document}

\title{FinBloom: Knowledge-Grounding Large Language Model with Real-Time Financial Data}

\author{\name Ankur Sinha \email asinha@iima.ac.in \\
       \addr Department of Operations and Decision Sciences\\
       Indian Institute of Management Ahmedabad\\
       Gujarat, India 380015
       \AND
       \name Chaitanya Agarwal \email chaitanyaa@iima.ac.in \\
       \addr Brij Disa Centre for Data Science and AI\\
       Indian Institute of Management Ahmedabad\\
       Gujarat, India 380015
       \AND
       \name Pekka Malo \email pekka.malo@aalto.fi \\
       \addr Department of Information and Service Economy\\
       Aalto University School of Business\\
       Helsinki, Finland}

\editor{~}

\maketitle

\begin{abstract}
Large language models (LLMs) excel at generating human-like responses but often struggle with interactive tasks that require access to real-time information. This limitation poses challenges in finance, where models must access up-to-date information, such as recent news or price movements, to support decision-making. To address this, we introduce Financial Agent, a knowledge-grounding approach for LLMs to handle financial queries using real-time text and tabular data. Our contributions are threefold: First, we develop a Financial Context Dataset of over 50,000 financial queries paired with the required context. Second, we develop FinBloom 7B, a custom 7 billion parameter LLM, by \textcolor{black}{fine-tuning Bloom 7B} on 14 million financial news articles from Reuters and Deutsche Presse-Agentur (DPA), alongside a random sample of 25\% from 12 million Securities and Exchange Commission (SEC) filings. Third, we fine-tune FinBloom 7B using the Financial Context Dataset to serve as a Financial Agent. This agent generates relevant financial context, enabling efficient real-time data retrieval to answer user queries. By reducing latency and eliminating the need for users to manually provide accurate data, our approach significantly enhances the capability of LLMs to handle dynamic financial tasks. Our proposed approach makes real-time financial decisions, algorithmic trading and other related tasks streamlined, and is valuable in contexts with high-velocity data flows.

\end{abstract}
%The 7 then subsequently fine-tuned this LLM on our custom dataset allowing us to obtain the context required and subsequently extract the tabular as well as news data to answer the query. 
%We perform a user study to show how investors and analysts perceive the output of the proposed knowledge-grounded approach. 
%Our proposed approach would make real-time financial decisions, algorithmic trading and other related tasks much more streamlined for the users.

%\begin{keyword}
%Financial Large Language Model \sep Generative Pre-Trained Transformer \sep knowledge-grounding \sep Natural Language Processing (NLP)
%\end{keyword}

%\end{frontmatter}

%\maketitle

\section{Introduction}
\begin{figure}[t]
  \centering %{.46\textwidth}
  \includegraphics[width=.6\textwidth]{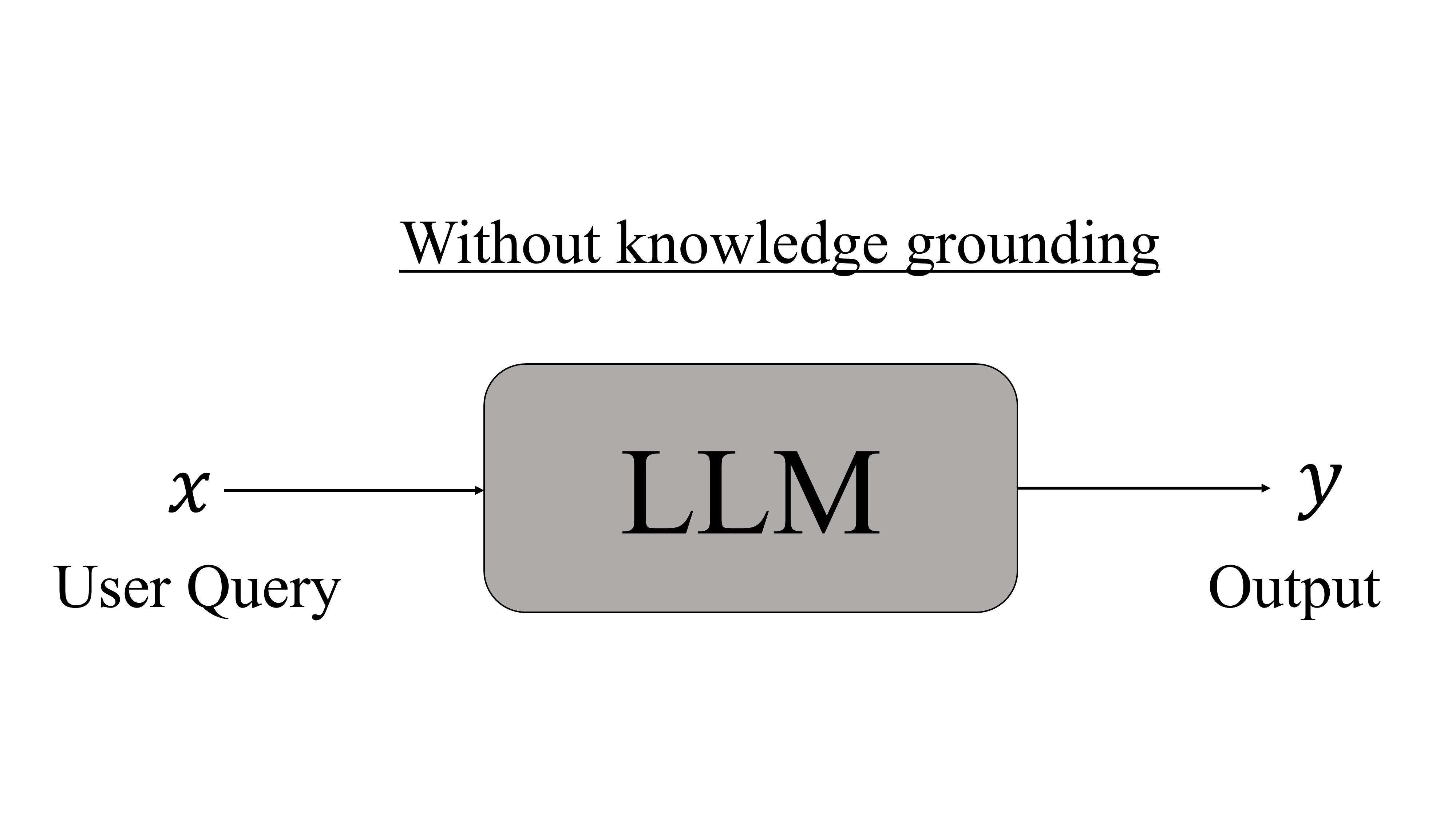}
  \captionsetup{aboveskip=-6pt, belowskip=0pt} % adjust spacing
  \caption{Static approach for answering user query without knowledge grounding}
  \label{fig:Financial Agent-withoutkg}
\end{figure}

Large Language Models (LLMs) use deep learning techniques to learn from massive text datasets to solve a variety of tasks using natural language \citep{devlin2018bert,brown2020language,scao2022bloom,zhang2022opt}. The most common tasks where these models have shown their proficiency include natural language generation, question-answering, reasoning, translation, summarization, etc. As the general-purpose models grow in size and learn from a larger data corpus, we observe that they become more versatile and develop the ability to handle multiple tasks that previously required separate specialized models. 
%However, we are still away from one-size-fits-all language models that can cater to the needs of all the areas. 
However, areas such as finance, healthcare, science, law, among others, often have their own domain-specific vocabulary that requires the fine-tuning of general-purpose models to perform domain-specific tasks \citep{wu2023bloomberggpt,luo2022biogpt,taylor2022galactica}. As LLMs get larger, it will be interesting to see whether general-purpose LLMs can work equally well as domain-specific LLMs simply with appropriate prompts. {\color{black}A rather more urgent problem that needs attention is that many of the domains require access to real-time data arising from several sources to handle various tasks with high accuracy and efficiency, which is currently beyond the capability of frozen LLMs that are not grounded in real-time knowledge and work as static systems as shown in Figure~\ref{fig:Financial Agent-withoutkg}}. For instance, in the context of finance, a user is likely to prefer using an LLM that is knowledge-grounded with real-time news and asset prices as they would want to make decisions based on accurate financial news and tabular data related to their query. Unless one uses a financial LLM for educational purposes, it will be of limited use, as most of its responses will be based on stale data that is of little value to investors or analysts. However, there are innumerable challenges when it comes to training or fine-tuning an LLM on real-time data that we list below:
\begin{enumerate}
\item Training or fine-tuning an LLM is very costly: Financial data such as asset prices change with a high frequency and similarly news data with an ability to impact asset prices is generated continuously in various parts of the world. Training or fine-tuning LLMs require an update in the model parameters which is quite time-consuming. Even if one attempts a small duration of fine-tuning based on batches of most recent numeric and text data, the LLM will still remain outdated as financial data flows at high velocity. Unless LLMs have the capability of quick online learning, knowledge-grounding based on frequent training or fine-tuning will not entirely help.
\item Lack of knowledge on what the LLM has learned: Most LLMs learn from massive data corpus, but as users of the system, we do not know what the model has learned and where it may commit errors. Only intensive testing of the models often reveal its weakness. Therefore, without a complete knowledge that LLM has learned from the new data appropriately, one may not put it to use in financial contexts where decisions have to be made based on new data.
\end{enumerate}
Handling the above challenges would be useful for any general-purpose or domain-specific LLM; therefore, there is a large body of recent research on knowledge-grounding of LLMs \citep{li2021knowledge,peng2023check,carta2023grounding}. An alternative approach to solve the above problem is to keep that LLM frozen, that is, do not change or fine-tune the weights of the LLM, rather add modules to the LLM for knowledge-grounding. \textcolor{black}{\cite{zhang2024multimodal} implemented this method of knowledge grounding in the financial domain, however, their approach focuses specifically on making trading decisions by leveraging LLMs to capitalize on market intelligence and technical indicators. Our study takes a distinct approach by designing and implementing modules that enable retrieval and incorporation of pertinent financial data for answering any financial query. This data augmentation is intended to enhance the LLM's capacity to respond accurately to a diverse array of financial inquiries.} The first module is the Data Module that provides access to real-time information (numeric and text data) that is flowing, and the second module consists of a Financial Agent that interprets the user query and extracts relevant data from the Data Module, thereby constructing the financial context required to answer the query. Advanced modules for video interpretation or audio analysis can also be added to the framework.

In this paper, we present a Financial Agent for grounding large language models in financial knowledge, enabling them to address financial queries effectively while managing dynamic text and tabular data streams. Our methodology leverages an external agent comprising a 7 billion parameter LLM integrated with real-time data sources to construct the contextual information required by a larger LLM (ChatGPT) for accurate query resolution. Our contributions in the paper are as follows:
\begin{enumerate}
\item We develop a custom Financial Context Dataset\footnote{Dataset: \url{https://www.kaggle.com/datasets/ankurzing/financial-context-dataset}}, containing over 50,000 financial queries paired with the corresponding contextual information needed for the resolution of the queries.
\item We fine-tune a 7 billion parameter LLM, referred to as FinBloom 7B\footnote{FinBloom 7B LLM: \url{https://huggingface.co/Chaitanya14/FinBloom_7B}}, on over 14 million financial news articles sourced from Reuters and Deutsche Presse-Agentur (DPA), equipping it with versatile capabilities for financial tasks. We also use a random sample of 25\% from 12 million documents from Securities and Exchange (SEC) filings along with news data.
\item We fine-tune FinBloom 7B on the Financial Context Dataset to create a Financial Agent\footnote{FinBloom 7B Agent: \url{https://huggingface.co/Chaitanya14/Financial_Agent}} that generates the context required to answer a user query and extracts data from a data module to effectively handle the financial query.
%\item The dataset and the models are being released publicly and can be accessed through the following links:
%\begin{enumerate}[leftmargin=0pt]
%\url{https://datasetAndModel.com}
%\end{enumerate}
\end{enumerate}
The proposed approach facilitates the extraction of high-velocity real-time financial data, enabling the effective resolution of user queries with support from the larger LLM (say GPT 3.5 or GPT 4). As a part of this study, we release the 50,000 Financial Context Dataset, 7 billion parameter domain-specific LLM and fine-tuned Financial Agent for context generation. We are unable to release the raw data, i.e. 14 million financial news articles from Reuters and DPA, because of contractual reasons. 
%We also release a package which integrates our contributions in this paper with other modules making the deployment of state-of-the-art artificial intelligence approaches simple for finance.

%To enable the Financial Agent to interpret the user's query and get relevant context to answer it, we built a custom dataset of over 50000 user queries and their corresponding contexts required to answer them, and then trained an LLM on this dataset which would take user's queries as inputs and then output the relevant context required to answer the query. Using the context provide by the LLM, relevant data can be extracted from the data module. The extracted data is then combined with the user query in the form of conversation history and then passed to the LLM. This allows the LLM to incorporate recent information while answering user queries.
\begin{figure}[t]
  \centering %{.46\textwidth}
  \includegraphics[width=.9\textwidth]{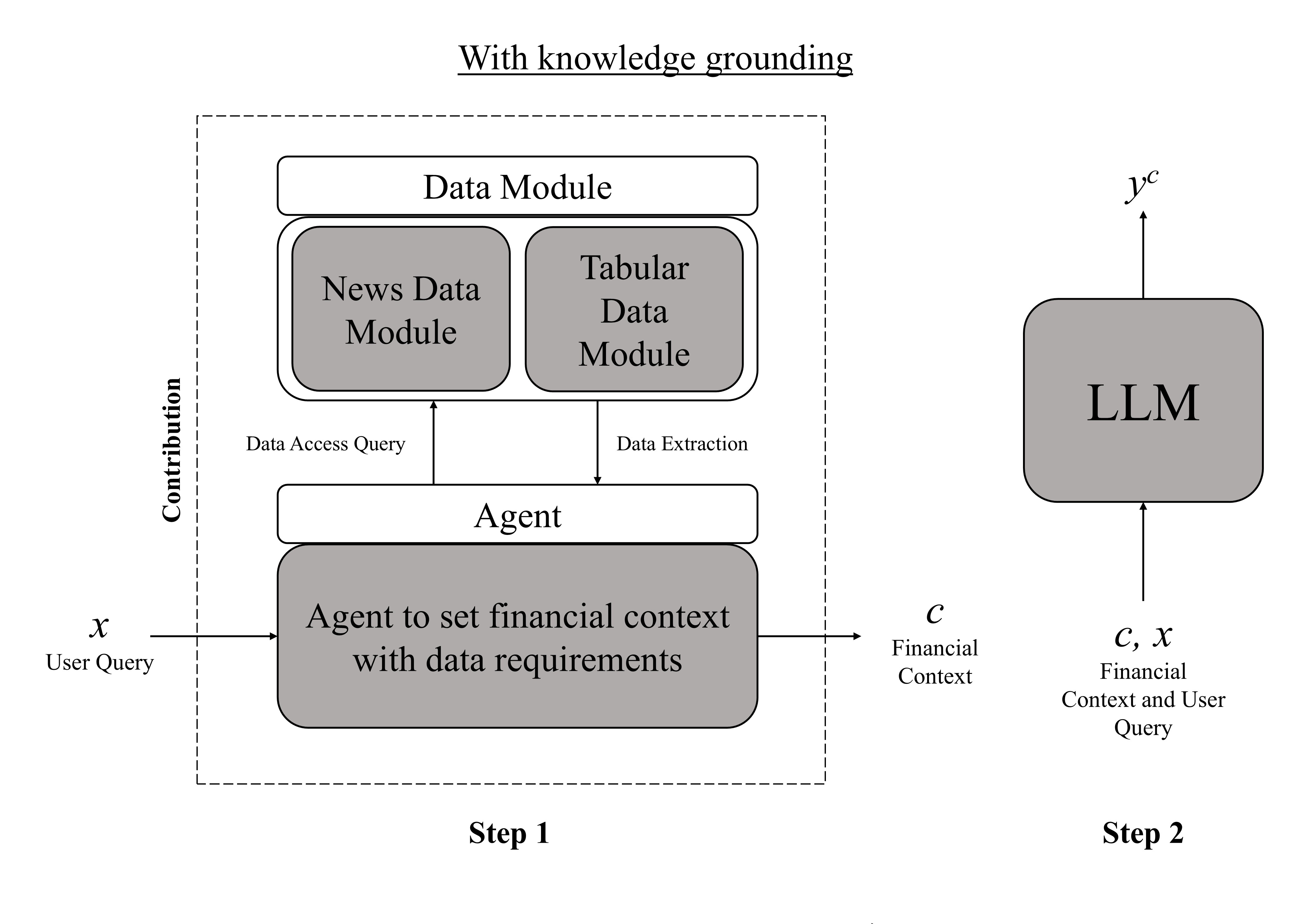}
  \captionsetup{aboveskip=-6pt, belowskip=0pt}
  \caption{Dynamic approach for answering user query with knowledge grounding. The context (relevant dynamic information) is generated using the Data Module and the Agent, which is one of the main contributions of the paper.}
  \label{fig:Financial Agent-withkg}
\end{figure}
Existing approaches to knowledge-grounding in LLMs follow different methodologies, which have a number of similarities, but are known in the literature with different terminologies. We address these approaches in the later part of the paper. We adopt an agent-based procedure to perform knowledge-grounding keeping in mind that financial data is high volume and high velocity for which high latency approaches may not be suitable.
{\color{black}
An overview of the proposed system---designed to respond efficiently to dynamic, real-time financial information---is shown in Figure~\ref{fig:Financial Agent-withkg}. 
}
The pipeline consists of three main components: a Data Module, a Financial Agent, and an LLM.
Because real-time financial information is typically available in tabular and textual forms, the Data Module maintains two continuously updated repositories: (i) a tabular database containing prices, financial metrics, and financial statements, and (ii) a text database that stores the latest news. Together, these repositories provide streaming context that is accessible to the Financial Agent. Given a user query \(x\), the Financial Agent first interprets the request and retrieves the most relevant news and tabular records from the Data Module. The retrieved content is then transformed into a textual context \(c\) and concatenated with the original query. The resulting input \((c, x)\) is provided to the LLM, enabling the model to generate an answer grounded in the retrieved, up-to-date financial data.
%\textcolor{black}{An efficient system that responds to dynamic and real-time information is proposed in this paper and illustrated in Figure~\ref{fig:Financial Agent-withkg}}. The complete pipeline has three major components, Data Module, Agent and an LLM. In the financial domain, real-time data is frequently presented in tabular or textual formats. To accommodate this, we maintain two repositories with streaming data: a tabular database containing up-to-date prices, financial metrics, financial statements, and a text database storing the latest news. These repositories collectively form the data module, which is accessible to the Financial Agent. 
%When a user query ($x$) is received, the Financial Agent analyzes the query and retrieves relevant news and tabular data from the data module. The retrieved data is then converted into a text format ($c$) and appended to the user query ($x$). This combined input ($c, x$) is passed to the LLM as part of a conversation, enabling it to generate an informed response that incorporates the contextualized real-time data. 
\textcolor{black}{Details about the Data Module implementation and retrieval are provided in Appendix~\ref{app:data-module}.}
\textcolor{black}{A detailed example on the working of the proposed system is provided in Appendix~\ref{appendix:example}. Additional examples in Appendix~\ref{appendix:example} also highlight how advanced LLMs may be of limited use to experts if not grounded with the necessary data.}

\begin{figure}[t]
  \centering %{.46\textwidth}
  \includegraphics[width=15 cm]{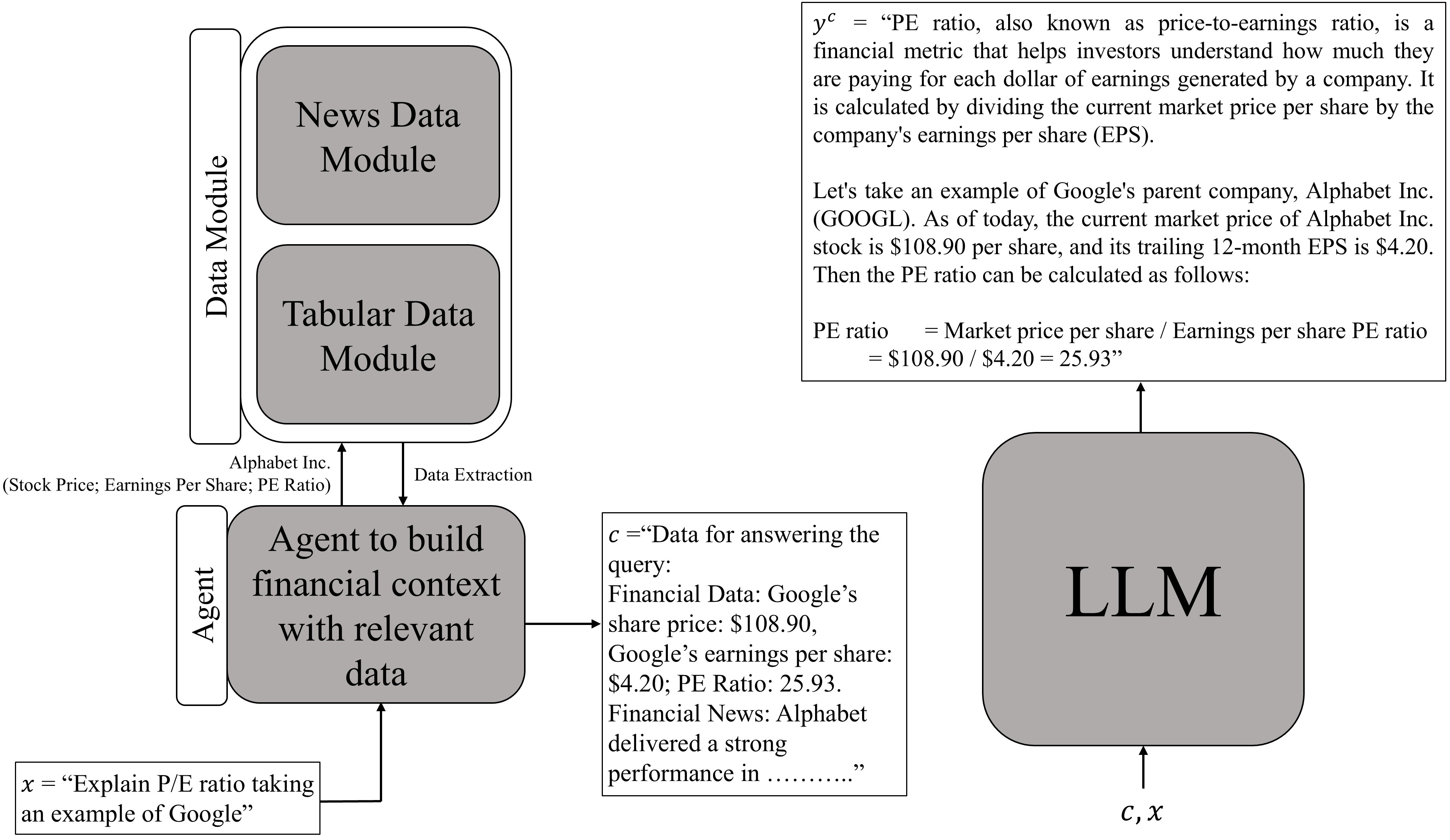}
  \caption{\textcolor{black}{Working of the proposed architecture: An example}}
  \label{fig:Financial Agent-exmp}
\end{figure}

A simple example of the operation of the proposed architecture is provided in Figure~\ref{fig:Financial Agent-exmp}. The example in the figure demonstrates how the Financial Agent leverages knowledge-grounding to process and respond to a financial query. When the user inputs the query ``Explain P/E ratio taking an example of Google,'' the Financial Agent performs contextualization by accessing the Data Module, which contains real-time information from tabular and textual repositories. Relevant data, such as Google’s share price (\$108.90) and earnings per share (EPS) (\$4.20), is extracted from the Data Module. This information is converted into a textual representation to create a context ($c$) that complements the original query. The query enrichment process then appends this context to the user query, forming a consolidated input ($c, x$) for the LLM. The enriched input enables the LLM to generate a detailed response, such as an explanation of the P/E ratio, its significance, and a calculation example using the provided data. By integrating real-time data extraction, contextualization, and query enrichment, the Financial Agent ensures precise and contextually relevant answers, demonstrating its capability for effective knowledge-grounding in dynamic financial scenarios.

%The existing approaches for knowledge-grounding of LLMs follow a similar procedure; however, researchers choose to refer to these modules with different names. In this paper, we perform knowledge-grounding using the stated procedure for our LLM. Note that in the context of finance, the real-time data is often in a tabular or text form. We maintain a repository of recent prices and financials in a tabular database, and a repository of recent news in a text database. This module, referred to as the data module, is accessible to the Financial Agent. Whenever, a user query ($x$) arrives, the Agent analyses the query and extracts news and tabular data ($d$) from the data module that is relevant to provide an appropriate answer to the user query. The extracted output is then converted into the form of text ($c (d)$), and is then appended to the user query ($x$). The appended tuple ($c (d),x$) is then passed to the LLM in the form of a conversation, which then answers the query by taking the relevant data into account. The architecture is explained through Figure~\ref{fig:Financial Agent-arch} and an example is provided through Figure~\ref{fig:Financial Agent-exmp}.

The structure of this paper is as follows. Section~\ref{sec:kg-survey} presents a comprehensive survey of recent advancements in knowledge-grounding of LLMs and financial datasets. Section~\ref{sec:fincontextdataset} provides details on the Financial Context Dataset consisting of 50,000 queries followed by Section~\ref{sec:finbloom} where we introduce a domain-specific LLM, referred to as FinBloom 7B. An extensive comparison of FinBloom 7B is performed against other popular models. In Section~\ref{sec:FinancialAgent}, we fine-tune FinBloom 7B on the Financial Context Dataset and create a Financial Agent capable of context generation and data retrieval. We also discuss its working along with the Data Module in our proposed knowledge-grounding framework in Section~\ref{sec:workflow}. Finally, conclusions and future directions for research are discussed in Section~\ref{sec:conclusions}.

%and Financial Agent, a domain-specific LLM for finance designed to leverage real-time data for effective knowledge-grounding. fine-tuning and experimental results are presented in Section~\ref{sec:results}, showcasing the performance and capabilities of the proposed system. 

%The paper is structured as follows. To begin with, we survey the recent work on knowledge-grounding of LLMs and financial datasets in Section~\ref{sec:kg-survey}. Thereafter, in Section~\ref{sec:Financial Agent} we discuss the architecture of Financial Agent in detail, which is a domain-specific LLM for finance that is knowledge-grounded with real-time data. We then provide the results from our experiments in Section~\ref{sec:results}. Finally, we conclude in Section~\ref{sec:conclusions}.

\section{Literature Review}\label{sec:kg-survey}
In this section, we provide a review on knowledge-grounding, contributions of LLMs in finance, and popular financial datasets for financial text mining. We also talk about the limitations of some of the popular knowledge-grounding approaches when applied to the area of finance.

\subsection{Knowledge-Grounding of LLMs}
The usual strategy for broadening the applicability of an LLM to a particular downstream task has conventionally entailed refining the model's parameters. This involves the process of training certain or all layers of the LLM on a bespoke dataset tailored to the specific requirements of the given downstream task \citep{radford2019language}. However, these strategies exhibit a considerable computational cost, especially if the data is dynamic and the model has to be updated frequently. %Moreover, they lead to the proliferation of modified model copies, each dedicated to a distinct downstream task, resulting in substantial storage demands. Given the large dimensions of contemporary LLMs, this approach becomes markedly impractical from both computational and resource allocation perspectives.
The fundamental rationale for knowledge-grounding in LLMs resides in the motivation to equip these models with the faculties of reasoning and contextual comprehension based on relevant external data, as opposed to them being regarded as repositories of static knowledge.

In recent years, a plethora of viable alternatives have emerged, offering potential enhancements to the conventional fine-tuning methodology.
\cite{wei2022chain} introduced ``chain-of-thought'' (CoT), a few-shot prompting technique for LLMs. CoT employs sequential examples within a prompt, comprising task inputs and intermediate steps, to drive contextual understanding and reasoning in large models. \cite{brown2020language} demonstrated GPT-3's ability to learn complex tasks in a few-shot setting.
But few-shot prompting requires manual effort to design the optimal prompt for any required downstream task. \cite{reynolds2021prompt} argued that few-shot learning isn't a method of task learning but rather a method of task location in the existing space of the model's learned tasks. They introduced the concept of metaprompt programming, through which the job of writing task-specific prompts can be assigned to the LLM itself. \cite{gao2020making} proposed an improvement in the few-shot strategy, where rather than relying on the arbitrary selection of random examples and their inclusion in the query, which lacks a guarantee of emphasizing the most informative demonstrations, their approach involved sequential random sampling of a single example from each class for every input resulting in the creation of multiple concise demonstration sets. \cite{shin2020autoprompt} introduced AUTOPROMPT, an approach that automatically constructs prompts by combining primary task inputs with an array of trigger tokens, adhering to a predetermined template. The set of trigger tokens used is same for all the inputs and is learned through a specialized adaptation of the gradient-based search strategy. The composite prompt is then supplied as input to a Mask Language Model. It was observed that for Natural Language Inference (NLI) tasks, AUTOPROMPT was comparable to a supervised fine-tuned BERT model.

Another limitation inherent in the few-shot prompting methodology stems from the constrained token capacity that the LLMs can intake. This might have thousands of examples that we need the LLM to learn from. A number of parameter efficient fine-tuning methodologies have been proposed to tackle this issue. 
\cite{liu2021gpt} proposed P-Tuning, a technique where the prompt tokens in the input embedding (containing context tokens and prompt tokens) are treated as pseudo tokens and mapped as trainable embedding tensors. This continuous prompt is modeled using a prompt encoder consisting of a bidirectional LSTM and is then optimized using the downstream loss function. Another unique approach is Prefix Tuning \citep{li2021prefix}, where a sequence of continuous task-specific vectors are prepended to the input of an autoregressive LM, or to both the encoder and decoder layers of a Encoder-Decoder Model. The prefix consists of trainable parameters which do not correspond to real tokens in a model's embedding. Instead of training the model's parameters on the loss function, only the parameters in the prefix are optimized. A similar methodology was used in Prompt Tuning \citep{lester2021power}, but without any intermediate-layer prefixes. \cite{dettmers2024qlora} proposed QLoRA, which introduces a memory-efficient fine-tuning method for LLMs by quantizing the pre-trained model to 4-bits and using Low Rank Adapters (LoRA), achieving comparable performance to 16-bit fine-tuning with reduced memory.

An alternate approach for the knowledge-grounding of LLMs involves equipping them with a retriever module that can access information relevant to the user's query from a database. \cite{lewis2020retrieval} introduced a retrieval-augmented generation (RAG) model, where they integrated a pre-trained retriever module with a pre-trained seq2seq model. The retriever provides latent documents conditioned on the input, and the {\it seq2seq} model then conditions on these latent documents together with the input to generate the output. \cite{dinan2018wizard} designed a dialogue model, where within the conversational framework, the chatbot is equipped with access to a curated collection of passages that maintain relevance to the ongoing discourse. At each turn, using a standard information retrieval system, the chatbot retrieves the top 7 articles for the last two turns of conversation. An attention mechanism is used to perform refined selection of specific sentences that will be used to create the next response. 
\cite{izacard2022few} created ATLAS, a RAG Model which is capable of few-shot learning.
\cite{peng2023check} presented LLM-Augmenter, a module which, in addition to knowledge retrieval and prompt generation, checks the response generated by a fixed LLM for hallucinations. If the response is not correct, it generates a feedback message which is used to improve the prompt. This cycle continues until the response by the LLM is verified. \cite{li2021knowledge} used a three step data cleaning procedure to supply an LLM with relevant context: retrieving associated triples from a knowledge graph, finding related triples by computing cosine similarity between the triples and the query, further refining the choices by estimating the semantic similarity score. 

Despite significant advancements in knowledge-grounding of LLMs, efficient deployment of them for real-time processing of enormous data flowing at high frequency, such as financial data, remains a challenge. 

{\color{black}
\subsection{Limitations of Traditional RAG Models in Retrieving Financial Data}
Recent years have witnessed a surge in the development and application of Retrieval-Augmented Generation (RAG) systems. By combining the power of LLMs with information retrieval techniques, RAG systems have the potential to revolutionize various industries. However, despite their promise, these systems still face several limitations that hinder their widespread adoption, particularly in complex and regulated domains.

A major source of confusion in the literature is that the term \emph{RAG} is often used broadly to denote any \emph{retrieval + generation} pipeline. In this broader sense, many systems (including ours) can be considered RAG-like. In this section, however, we use ``traditional RAG'' to specifically refer to the dominant design pattern in which a large corpus is embedded into a dense vector space and retrieved via approximate nearest-neighbour (ANN) search over text chunks. The limitations discussed below target dense-embedding retrieval over textified content, which is particularly fragile when the underlying knowledge is structured and fast-flowing financial data.

One of the core challenges in such RAG systems lies in the effectiveness of the retrieval mechanism. As highlighted by \cite{gupta2024comprehensive}, while powerful, retrieval can struggle with ambiguous queries and niche knowledge domains. The reliance on dense vector representations can sometimes lead to the retrieval of irrelevant documents, which can negatively impact the quality of the generated response. \cite{cuconasu2024power} further emphasized this point, demonstrating that high-scoring retrieved documents that are not directly relevant to the query can degrade LLM effectiveness. Moreover, the quality of the generated output can be influenced by the number of retrieved passages, as outlined in \cite{jin2024long}; while increasing the number of passages can initially improve quality, it can eventually lead to a decline in performance due to distraction and context dilution.

When integrating knowledge graphs (KGs) into RAG systems, additional challenges arise in effectively utilizing structured information. \cite{agrawal2024mindful} identified several critical failure points in existing KG-based RAG methods, including insufficient focus on question intent and inadequate context gathering from KG facts. The real-world applications of retrieval-augmented systems, especially in expert domains, are often characterized by complex and nuanced requirements. As noted by \cite{zhao2024retrieval}, the one-size-fits-all approach to data augmentation may not be suitable for all scenarios. In compliance-regulated sectors, systems must also adhere to stringent data privacy, security, and governance requirements; \cite{bruckhaus2024rag} highlighted the importance of ensuring that sensitive data is not exposed or misused during the retrieval and generation process.

%\paragraph{Why structured financial data is especially challenging for dense-vector RAG.}
Dealing with financial data presents additional challenges to traditional dense-vector RAG pipelines, primarily because the underlying knowledge is predominantly \emph{structured} (tables, time series, and numeric metrics), while dense-vector RAG is optimized for \emph{unstructured text}. Specifically:
\begin{enumerate}
    \item \textbf{Difficulties in representing financial data as retrievable text.} Financial data is predominantly stored in structured, tabular formats (spreadsheets, relational databases, and time-series stores). Traditional dense-vector RAG commonly requires converting these tables into unstructured text chunks and embedding them for ANN search. This conversion is often impractical at scale, introduces redundancy, and can create errors or inconsistencies (e.g., mismatched units, stale snapshots, or fragmented rows). Further, because tabular structure encodes meaning through schema, keys, and alignment, naive textification may discard or obscure critical relationships needed for precise retrieval and reasoning.
    \item \textbf{Retriever precision requirements for financial queries.} Financial queries often specify companies, financial metrics and date ranges. Dense-vector retrieval over textified tables is not guaranteed to return the exact cell-level values required, and small retrieval errors can lead to materially incorrect answers. Incomplete retrieval yields missing evidence, while excessive retrieval burdens the generator with irrelevant context, amplifying hallucinations and misinterpretations.
    \item \textbf{High-velocity, high-volume updates.} Financial datasets are frequently updated (earnings releases, macroeconomic indicators, market data, filings), and dense-vector indices require periodic re-embedding and re-indexing to stay current. In high-velocity settings, this pipeline can become costly and latency-prone, increasing the risk of stale retrieval and outdated generation.
\end{enumerate}

%\paragraph{Implication for our design.}
Motivated by these limitations, our system avoids embedding the structured financial database into a vector space and does not rely on ANN search over textified tables. Instead, structured financial retrieval is performed via \emph{schema-aligned query prediction}: a small text-to-text Financial Agent generates a Structured Data Request, which is compiled into exact data request queries over the relational store, returning precise responses. Only the retrieved slice is then linearized into text for the LLM to perform explanation and synthesis. 
}%\color ends

\begin{table}[!h]
\footnotesize
    \centering
    \caption{Summary of Financial Datasets and Resources}
    \begin{tabular}{|p{0.15\textwidth}|p{0.6\textwidth}|p{0.15\textwidth}|}
        \hline
        \textbf{Resource} & \textbf{Description} & \textbf{Reference} \\ \hline
        SentiWordNet & Establishes a lexical resource based on WordNet, assigning sentiment scores (objective, positive, negative) to individual word senses (synsets). & \cite{baccianella2010sentiwordnet} \\ \hline
        Financial Phrase Bank & Contains almost 5,000 snippets of text, gathered from financial news and press releases about Finnish companies traded on the NASDAQ OMX Nordic Exchange. Each snippet is labeled as positive, negative, or neutral. & \cite{malo2014good} \\ \hline
        SenticNet & A dataset for understanding sentiment in financial texts, combining artificial intelligence and Semantic Web technologies to better recognize, interpret, and process opinions within this context. & \cite{cambria2012senticnet} \\ \hline
        SEntFiN 1.0 & Incorporates entity-sentiment annotation to address the challenge of headlines containing multiple entities with potentially conflicting sentiments. This human-annotated dataset of over 10,700 news headlines provides insights into sentiment variations depending on specific entities in financial news. & \cite{sinha2022sentfin} \\ \hline
        Trillion Dollar Words & A dataset of tokenized and annotated Federal Open Market Committee (FOMC) speeches, meeting minutes, and press conference transcripts. This resource aims to understand the influence of monetary policy on financial markets by analyzing the language used by central bank. & \cite{shah2023trillion} \\ \hline
        REFinD & An annotated dataset of relations within financial documents with approximately 29,000 instances and 22 relation types across eight entity pairs (e.g., person-title, org-money), extracted from SEC 10-X filings. Useful for identifying relationships between entities in financial reports. & \cite{kaur2023refind} \\ \hline
        Gold Commodity Dataset & Compiled from diverse news sources and evaluated by human experts, allowing researchers to explore how news headlines influence price movements, asset comparisons, and other financial events. & \cite{sinha2021impact} \\ \hline
        FiNER & A dataset designed for Named Entity Recognition (NER) within the financial landscape, aiding in identifying and classifying financial companies within text data. & \cite{shah2023finer} \\ \hline
        MULTIFIN & Provides a multilingual financial NLP dataset with real-world financial article headlines in 15 languages. Annotated with high-level and low-level topics, facilitating multi-class and multi-label classification tasks. & \cite{jorgensen2023multifin} \\ \hline
        FINQA & A collection of question-answer pairs with in-depth analysis of financial reports. Written by financial experts, it focuses on deep reasoning over financial data, providing training data for answering complex financial queries. & \cite{chen2021finqa} \\ \hline
    \end{tabular}
    \vspace{-5mm}
    \label{tab:financial-datasets}
\end{table}

\subsection{Financial Datasets Review}   
The domain of financial language processing has witnessed a significant rise in the development of annotated datasets, each serving distinct purposes within the field. Through Table~\ref{tab:financial-datasets}, we have highlighted some prominent examples that have been widely used in the area of finance for training models. In the next sections, we discuss the core contributions of the paper in detail. The key contribution is the creation of a Financial Agent that is capable of understanding the context needed for a user query, extracting the relevant data from the databases, and supporting an LLM to handle the contextualized query effectively.

\section{Financial Context Dataset}\label{sec:fincontextdataset}
For any given user query, a Financial Agent is expected to provide the relevant context, which includes numeric and text data, which can be appended to the user query for further processing. In this section, we discuss the first core contribution of the paper, which is the Financial Context Dataset, consisting of 50,000 samples of user query and its corresponding context. The proposed dataset can be used to fine-tune a Financial Agent specifically for context identification and data extraction. In this section, we present a detailed account of the dataset creation and its description.

\begin{figure}[t]
  \centering 
  \includegraphics[width=15cm]{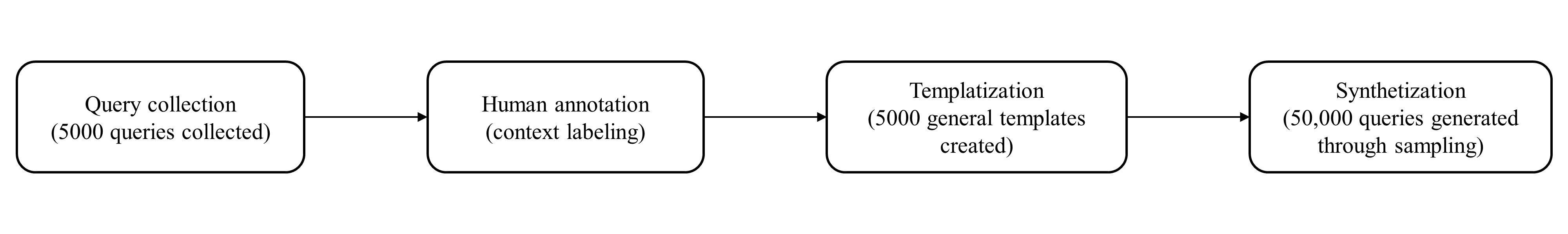}
  \caption{\textcolor{black}{Steps for dataset generation}}
  \label{fig:dataset-flow}
\end{figure}

To fine-tune our model effectively, the dataset should be diverse in terms of sentence structure and comprehensive in terms of financial metrics and companies. Specifically, the dataset should include queries in a variety of forms, such as simple yes/no questions, comparative questions, and open-ended questions. This will help the model to learn to understand the different ways that users can present their queries. Additionally, the dataset should include a large number of financial metrics and companies, in the form of their many possible names and other ways they are referred to generally. This diversity in the dataset will enable our model to generalize well to a wide range of user queries, ensuring that it can accurately identify the required information. To construct this dataset, which consists of natural language user queries and their corresponding structured data requests, we followed a multi-method query sampling approach. \textcolor{black}{The first method involved collecting queries sent by retail customers over technology channels to a brokerage and investment firm, like email, web-based forms, queries on app and Whatsapp.} The second method involved interaction and consultation with 10 financial advisors on the kind of queries retail investors commonly have. The third method involved interviews with 20 retail investors about common questions that come to their mind before investing in a particular stock or forming a portfolio. \textcolor{black}{Details of these interactions are provided in Appendix~\ref{sec:templateApp}.} 
Based on the three methods of data collection; i.e. technology channel queries, advisor interviews and investor interviews; we created 5000 query templates, in which 4000 templates were extracted from technology channels, 500 were created with financial advisors, and another 500 were created with retail investors. This approach ensured the incorporation of a broad scope of financial queries, resulting in a diverse dataset. In order to extend the dataset to 50000 queries, the 5000 templates were scaled-up by randomly varying various aspects within each template. Table~\ref{tab:template} shows 4 queries and the corresponding template that we created. Once the template is created, it is possible to map various kinds of queries to each template, for instance, Table~\ref{tab:template2} in Appendix~\ref{sec:templateApp} provides randomization of various aspects in the template leading to a larger set of queries in our dataset. \textcolor{black}{The flow for the dataset generation process is provided through Figure \ref{fig:dataset-flow}.}

\begin{table}[t]
\centering
\caption{Example queries and their corresponding templates.}
\begin{tabular}{|p{0.45\textwidth}|p{0.45\textwidth}|}
\hline
\textbf{Query} & \textbf{Template} \\ \hline
Judging by Apple's revenue and profit margins in Q4 2024, is it a good investment opportunity? & \textit{Judging by [company]'s [metrics][date], is it a good investment opportunity?} \\ \hline
Evaluate the return-on-equity ratio associated with Infosys in 2022. & \textit{Evaluate the [metrics] associated with [company][date].} \\ \hline
List the top 5 companies in Q3 2023 in terms of stock price growth. & \textit{List the top 5 companies [date] in terms of [metrics].} \\ \hline
Were there any improvements seen in the market share of Samsung in 2023 as compared to Apple? & \textit{Were there any improvements seen in the [metrics] of [company1][date] as compared to [company2]?} \\ \hline
\end{tabular}
\label{tab:template}
\end{table}

It is obvious that not all financial metrics or companies can be mapped to each of the queries because of the differences in how the queries are structured. For example, the metrics `revenue breakdown', `market segmentation', or `customer demographics' can be mapped to the template, \textit{Provide details about the [metrics] of companies [date].}, but not to the template, \textit{Did the [metrics] of [companies] increase [date]?} because they may not inherently be quantitative or trackable over time. 
%Descriptive metrics, such as revenue breakdown, customer demographics, and market segmentation, are well-suited to templates like \textit{Provide details about the [metrics] of companies [date]}, as they allow for detailed elaboration. Conversely, metrics that exhibit clear numerical trends over time, such as profit margins, sales growth, or stock prices, are better suited to templates like \textit{Did the [metrics] of [companies] increase [date]?}, which require measurable changes or directional insights. This classification ensures that each metric is mapped appropriately, preventing mismatches and enhancing the relevance of the generated queries.
Metrics that exhibit clear numerical trends over time, such as profit margins, sales growth, or stock prices, are better suited to templates like \textit{Did the [metrics] of [companies] increase [date]?}, which require measurable changes or directional insights. Therefore, the metric set maintained by us is carefully curated to align with the specific structure and intent of each template. 
%Our approach ensures that each metric is mapped appropriately, preventing mismatches and enhancing the relevance of the generated queries.

We maintain a comprehensive repository of financial metrics \texttt{[metrics]}, company names \texttt{[company]}, industries \texttt{[industry]}, and augment it with time range variations \texttt{[date]} and numeric variations \texttt{[number]} to systematically generate a diverse set of natural language queries. This template-driven approach to query generation ensures that the resulting dataset is both highly diverse and semantically accurate. By leveraging predefined templates, we produce a dataset that encompasses a broad spectrum of financial metrics and company-specific contexts. %significantly surpassing the diversity and informativeness of datasets created through purely randomized query generation. 
Furthermore, this approach eliminates the need for manual extraction and labeling of financial information, thereby enhancing both efficiency and scalability in dataset creation.

%\begin PLACEHOLDER for column 2 and 3 description
After creating 50000 queries using a template-based approach, we create two additional columns in the dataset, namely, ``Required Data'' and ``Structured Data Request''. The ``Required Data'' column specifies the necessary information required for answering the query, comprising company names, their specific metrics required for analysis, and the corresponding date ranges for data retrieval. Each metric is associated with a list of ``Related Metrics'', which are also to be retrieved from the Data Module. A snippet of our dataset, 3 out of 50000 rows, with all three columns is shown in Table~\ref{tab:snippet}. The inclusion of related metrics serves four primary purposes: (1) It provides additional contextual information for an efficient processing of the query by an LLM. (2) It enhances the robustness of the analysis by mitigating potential ambiguities or inaccuracies in the data. Financial metrics often have complex interdependencies, and providing related metrics ensures that even if the primary metric is misunderstood, incomplete, or unavailable, the LLM can derive or approximate insights using the supplementary data. This redundancy in data retrieval acts as a safeguard against potential errors, thereby improving the reliability of the query responses. (3) It enables the generation of enriched and multidimensional insights that go beyond the scope of the original query. By retrieving additional metrics linked to the primary ones, the system can identify trends, correlations, or anomalies that might not be apparent when considering the primary metric alone. This capability allows the LLM to provide more comprehensive, informed, and actionable responses, thereby increasing the value of the analysis for end-users. (4) By leveraging ``Related Metrics'', the LLM can perform multi-metric calculations (e.g., ratios, averages, or differences), which can be useful for educational queries or may also lead to insights for further interactions. For instance, see Figure~\ref{fig:Financial Agent-exmp}, where the query is about the P / E ratio of a company, but it can be effectively tackled only when the data on the ``share price'' and the ``earnings per share'' are known.

\begin{landscape}
\begin{table}[h!]
\begin{center}
 \begin{tabular}{|p{7cm}|p{7cm}|p{7cm}|}
 \hline
 \textbf{Query} & \textbf{Required Data} & \textbf{Structured Data Request}\\
 \hline
 Based on Amcor's Acid Test Ratio, Bid Size and Cash Conversion Efficiency Ratio for the previous 6 months, should I invest in it? & Companies: Amcor plc \newline Metrics: Quick Ratio (Related Metrics: Cash, Cash Equivalents, Marketable Securities, Accounts Receivable, Current Liabilities); Bid Size (Related Metrics: Quantity of shares, Multiple bid prices, Depth of the market, Order book); Cash Conversion Efficiency Ratio (Related Metrics: Cash flow from operations, Net income)\newline Dates: for the previous 6 months & (Amcor plc) \newline (Quick Ratio; Cash; Cash Equivalents; Marketable Securities; Accounts Receivable; Current Liabilities; Bid Size; Quantity of shares; Multiple bid prices; Depth of the market; Order book; Cash Conversion Efficiency Ratio; Cash flow from operations; Net income)\newline (7/1/2024 - 7/7/2024)\\
 \hline
 Give an overview of adobe and its competitor's Sep 2018 sales revenue, EVA. & Companies: Adobe Inc.; Adobe Inc. Peers\newline
Metrics: Sales Revenue (Related Metrics: Total Revenue); Economic Value Added (Related Metrics: Net operating profit after tax (NOPAT), Cost of capital)\newline
Dates: Sep 2018 &  (Adobe Inc.; Adobe Inc. Peers)\newline
 (Sales Revenue; Total Revenue; Economic Value Added; Net operating profit after tax (NOPAT); Cost of capital)\newline
 (1/9/2018 - 30/9/2018)\\
\hline
What were the return on average assets and ROWC of Halliburton co. and other companies in the energy sector from Apr 2016 to Jul 2017 compared to their Gross Profit Margin and CROAFA? & Companies: Halliburton Co.; energy Companies\newline
Metrics: Return on Average Assets (Related Metrics: Net income, Average total assets); Return on Working Capital (Related Metrics: Net income, Working capital); Gross Profit Margin (Related Metrics: Revenue, Cost of Goods Sold (COGS)); Cash Return on Average Fixed Assets (Related Metrics: Operating cash flow, Average fixed assets)\newline
Dates: from Apr 2016 to Jul 2017 & (Halliburton Co.; Energy Companies)\newline (Return on Average Assets; Net income; Average total assets; Return on Working Capital; Net income; Working capital; Gross Profit Margin; Revenue; Cost of Goods Sold (COGS); Cash Return on Average Fixed Assets; Operating cash flow; Average fixed assets)\newline
 (1/4/2016 - 1/7/2017)\\
\hline
\end{tabular}
\captionsetup{justification=centering}
\caption{Table showing snippet of the proposed dataset containing financial queries and their corresponding data request columns. A set of 3 out of 50000 rows are shown.}
\label{tab:snippet}
\end{center}
\end{table}
\end{landscape}

We have closely studied all the financial metrics that occur in financial text and generated a predefined mapping that links each financial metric to a set of related metrics. Identification and selection of these related metrics were carried out through a meticulous human-driven process to ensure accuracy and domain relevance. While modern techniques, such as computing cosine similarity using transformer-based models, could have been employed to identify related metrics automatically, we prioritized a human-based approach to maintain correctness and contextual appropriateness. This effort ensures a robust and reliable dataset, ensuring that the relationships between metrics are meaningful and aligned with practical financial analysis.
Finally, the ``Structured Data Request'' field is prepared that represents a formalized, machine-readable version of the ``Required Data'' field, designed for direct input into the Data Module for efficient data retrieval. 

In a later section, we will utilize the prepared dataset for fine-tuning an LLM-based Financial Agent for identifying the relevant data required for answering the query. Before the fine-tuning exercise, in the next section we first create a financial LLM that can be fine-tuned for any downstream task.

\section{FinBloom 7B: A Large Language Model for Finance}\label{sec:finbloom}
Our approach for addressing user queries requires a Financial Agent for forming a structured request of the financial data that the Data Module can understand and respond to. The user's query is typically in natural language, which means that it may not explicitly contain all of the necessary information, like recent numbers and news, that is needed to answer it. %Additionally, the data that is needed may not be present in the query in a format that the Data Module can interpret. 
Hence, the Agent needs to analyze the user's query and intelligently identify the essential financial metrics, as well as any related metrics and news, that are needed to answer the query. It also needs to identify the companies for which the data is required, as well as the date ranges for which the data is needed. It finally needs to ensure that these requirements are presented in the form of a data request function that the Data Module can interpret. Manually specifying a set of hard rules to handle all possible user queries and perform these tasks would be very tedious and complicated. Therefore, a better approach is to fine-tune a text-to-text generation model on a large dataset of natural language user queries for generating the corresponding structured query formats which contain all the information required to answer the user query. There are quite a few financial LLMs such as FiMA (\cite{xie2023pixiu}), FinGPT (\cite{yang2023fingpt}), CFGPT (\cite{li2023cfgpt}), InvestLM (\cite{yang2023investlmlargelanguagemodel}) available that can be fine-tuned to serve as an agent for the task of creating structured data request. However, most of these LLMs exhibit limitations that may hinder their effectiveness in real-world applications, especially for the purpose of identifying financial context for a given query.

Many domain-specific models are typically developed by starting with a general-purpose base model, which is subsequently fine-tuned using human-curated financial datasets (for example, FinMA (\cite{xie2023pixiu}), InvestLM (\cite{yang2023investlmlargelanguagemodel})) such as instruction datasets, question-answer datasets, structured datasets, or synthetic datasets. To the best of our knowledge, most existing financial LLMs are not trained on high-quality, large-scale domain-specific text corpora. This limitation primarily arises from the scarcity of extensive and ethically\footnote{\textcolor{black}{Ethical use in the context of financial LLM training involves ensuring that the text corpora used for model development are obtained with proper permissions and comply with copyright, privacy, and data ownership regulations. This includes avoiding the use of proprietary data without authorization and adhering to fair-use policies when sourcing publicly available content. All data used in this study is ethically sourced with appropriate authorization.}} usable financial text corpora, as such resources are often difficult to access or compile from online channels. Many human-curated financial datasets used for training finance-specific LLMs are of limited size and highly structured; therefore, it may fail to convey domain-specific nuances and may not capture the complex inter-dependencies inherent in financial metrics and text. Human-curated datasets are also often inherent with human biases, leading to LLMs with shallow understanding of financial text. Using financial text corpora, such as large volumes of written financial text (books, news, filings, etc.), enriches the model’s knowledge of domain-relevant terminology, concepts, and context. Even in cases where one desires to train domain-specific instruction models, it is advisable to take a hybrid approach involving training on text-corpora followed by fine-tuning on instruction dataset. The existing domain-specific LLMs lack this phase of training on financial text corpora. Some financial LLMs such as FinGPT (\cite{yang2023fingpt}), CFGPT (\cite{li2023cfgpt}) are trained on publicly available financial text corpora, but are not competitive when evaluated on financial benchmarks in the later part of this section.

To create the financial LLM, we choose a foundational model with robust general language understanding and the potential for domain-specific specialization in finance. We began by selecting the Bloom 7B parameter model as the base, leveraging its strong linguistic capabilities as a foundation for fine-tuning. To equip the model with financial expertise, we used a large corpus of financial news articles procured from Reuters and Deutsche Presse-Agentur (DPA). This dataset encompasses a wide array of financial topics, including market trends, economic indicators, corporate developments, and regulatory changes. The Reuters dataset spans the period from January 1, 2003, to December 31, 2012, while the DPA dataset covers the timeframe from June 1, 2001, to May 31, 2011. The combined dataset consists of over 14 million articles.
The high-quality financial data used for training is expected to lead to an LLM that understands interactions between various financial metrics and concepts well. We also utilized the publicly available SEC filings data from the period 31st Mar 2009 to 31st Oct 2023. The SEC filings consisted of overall 12.23 million documents but we utilized a random sample of 25\% for the purpose of fine-tuning to avoid biasing the model with a large volume of ``regulatory reporting data''. The details of the three datasets are provided in Table~\ref{tab:dataset-details}.
\begin{table}[h!]
\small
\centering
\begin{tabular}{|l|r|r|r|l|}
\hline
\textbf{Dataset} & \textbf{Documents} & \textbf{Mean Words} & \textbf{Mean Tokens} & \textbf{Time Period} \\ \hline
Reuters & 14,574,641   & 369.23        & 459.09             & 1st Jan 2003-31st Dec 2012                     \\ \hline
DPA     & 387,187   & 286.20        & 390.37             & 1st Jun 2001-31st May 2011                     \\ \hline
SEC     & 12,238,570 & 379.96       & 536.56             & 31st Mar 2009-31st Oct 2023                    \\ \hline
\end{tabular}
\caption{Dataset statistics with with mean words and mean tokens per document.}
\label{tab:dataset-details}
\end{table}
fine-tuning was performed using the efficient QLoRA method (\cite{dettmers2024qlora}), which facilitated significant performance improvements with minimal computational resources. The \textcolor{black}{fine-tuning} process spanned four epochs on a Tesla T4 GPU, requiring over 2000 hours. The model derived from this training will be referred to as ``FinBloom 7B'' in the later parts of the paper.

To evaluate the capabilities of FinBloom 7B across a diverse range of financial tasks, we adopted the FinBen benchmark (\cite{xie2024finben}). This comprehensive benchmark consists of multiple datasets, each assessing specific competencies such as information extraction, textual analysis, question answering, text generation, forecasting, risk management, and decision-making. Details of the FinBen benchmark are provided in Appendix~\ref{sec:finben}. By subjecting our model to this rigorous evaluation, we aimed to measure its effectiveness in handling various real-world financial applications. Table~\ref{table:comparison1} presents the performance comparison of FinBloom 7B against other widely used LLMs (FinMA-7B ( \cite{xie2023pixiu}), FinGPT 7b-lora (\cite{yang2023fingpt}), CFGPT sft-7B-Full (\cite{li2023cfgpt}) across the FinBen benchmark, providing insights into its relative strengths and specialization within the financial domain. \textcolor{black}{There are 35 datasets in the FinBen benchmark out of which we first consider 25 datasets for which one can compute the F1 score or its variation.} \textcolor{black}{The results reported for FinBloom 7B are without supervised fine-tuning on the datasets.} On these 25 datasets, FinBloom 7B shows the best average performance. It is interesting to note that FinBloom 7B has not seen any of the data in Table~\ref{table:comparison1} during the training phase. FinMA 7B, that is a close competitor to FinBloom 7B has seen significant amount of data (4 out of 25 datasets) during the training phase. {\color{black}Several models achieved scores of 0.00 or performed poorly, primarily due to limitations such as inadequate instruction tuning, which led to failure in following task-specific guidelines, or the generation of hallucinated responses and outputs in incorrect formats. These issues were particularly pronounced in information extraction tasks, including named entity recognition, causal detection, numeric labeling, and textual analogy parsing.}
%Models with a smaller number of parameters demonstrated significant difficulty with these tasks, especially in the absence of fine-tuning on the relevant datasets.

We have provided a comparison of all these models against popular LLMs like ChatGPT, GPT 4, Gemini and LLaMA on FinBen benchmark in Table~\ref{tab:allComparison} in Appendix~\ref{sec:finben}. Clearly GPT-4 is the best performer among all LLMs. Though the number of model parameters in GPT-4 is officially unknown, it is estimated to have more than a trillion model parameters. \textcolor{black}{Appendix~\ref{sec:finben} provides further details on the procedure that we used to evaluate the models across the datasets in FinBen.}

\begin{table}[h!]
\centering
\setlength{\arrayrulewidth}{0.1mm}
\setlength{\tabcolsep}{1pt}
\renewcommand{\arraystretch}{1.8}
\fontsize{6pt}{5pt}\selectfont
\begin{tabular} 
{ m{2.0cm} m{3.6cm} m{2.1cm} m{2.1cm} m{2.0cm} m{1.9cm} m{1.9cm} }
\hline
\textbf{Dataset} & \textbf{Task} & \textbf{Metrics} & \textbf{FinMA 7B} & \textbf{FinGPT \newline 7B-lora} & \textbf{CFGPT \newline sft-7B-Full} & \textbf{FinBloom 7B} \\
\hline
NER & named entity recognition & EntityF1 & \textbf{0.69} & 0.00 & 0.00 & 0.00\\
FINER-ORD & named entity recognition & EntityF1 & 0.00 & 0.00 & 0.00 & 0.00\\
FinRED & relation extraction & F1 & 0.00 & 0.00 & 0.00 & 0.00\\
SC & causal classification & F1 & 0.19 & 0.00 & 0.15 & \textbf{0.84} \\
CD & causal detection & F1 & 0.00 & 0.00 & 0.00 & 0.00 \\
FNXL & numeric labeling & EntityF1 & 0.00 & 0.00 & 0.00 & 0.00 \\
FSRL & textual analogy parsing & EntityF1 & 0.00 & 0.00 & 0.00 & 0.00 \\
\hline
FPB  & sentiment analysis & F1 & \textbf{0.88} & 0.00 & 0.35 & 0.32 \\
                      % & Acc & \textbf{0.88} & 0.00 & 0.26 & 0.31 \\
FiQA-SA & sentiment analysis & F1 & \textbf{0.79} & 0.00 &0.42 & 0.47 \\

Headlines & news headline classification & AvgF1 & \textbf{0.97} & 0.60 & 0.61 & 0.45 \\
FOMC  & hawkish-dovish classification & F1 & \textbf{0.49} & 0.00 & 0.16 & 0.20\\
                      % & Acc & 0.46 & 0.00 & 0.21 & 0.25 \\
FinArg-AUC & argument unit classification & MicroF1 & 0.27 & 0.00 & 0.05 & \textbf{0.81} \\
FinArg-ARC & argument relation classification & MicroF1 & 0.08 & 0.00 & 0.05 & \textbf{0.20} \\
MultiFin & multi-class classification & MicroF1 & 0.14 & 0.00 & 0.05 & \textbf{0.25} \\
MA & deal completeness classification & MicroF1 & 0.45 & 0.00 & 0.25 & \textbf{0.48} \\
MLESG & ESG Issue Identification & MicroF1 & 0.00 & 0.00 & 0.01 & \textbf{0.07} \\
\hline
{German}  & credit scoring & F1 & 0.17 & 0.52 & 0.53 & \textbf{0.53} \\
                      % & MCC & 0.00 & 0.00 & 0.00 & -0.04 \\
{Australian}  & credit scoring & F1 & 0.41 & 0.38 & 0.29 & \textbf{0.51} \\
                      % & MCC & 0.00 & 0.11 & -0.10 & 0.08 \\
{LendingClub}  & credit scoring & F1 & \textbf{0.61} & 0.00 & 0.05 & 0.53 \\
                      % & MCC & 0.00 & 0.00 & 0.01 & -0.02 \\
{ccf}  & fraud detection & F1 & 0.00 & \textbf{1.00} & 0.00 & 0.70 \\
                      % & MCC & 0.00 & 0.00 & 0.00 & -0.01 \\
{ccfraud}  & fraud detection & F1 & 0.01 & 0.00 & 0.03 & \textbf{0.77} \\
                      % & MCC & -0.06 & 0.00 & 0.01 & -0.01 \\
{polish}  & financial distress classification & F1 & \textbf{0.92} & 0.30 & 0.40 & 0.30 \\
                      % & MCC & -0.01 & 0.00 & -0.02 & 0.00 \\
{taiwan}  & financial distress classification & F1 & \textbf{0.95} & 0.60 & 0.70 & 0.32 \\
                      % & MCC & 0.00 & -0.02 & 0.00 & 0.02 \\
{portoseguro}  & claim analysis & F1 & 0.04 & \textbf{0.96} & 0.00 & 0.41 \\
                      % & MCC & \textbf{0.01} & 0.00 & 0.00 & 0.00 \\
{travelinsurance}  & claim analysis & F1 & 0.00 & \textbf{0.98} & 0.03 & 0.50 \\
                      % & MCC & 0.00 & 0.00 & 0.01 & 0.01 \\
\hline
\multicolumn{3}{c}{\textbf{Average F1 Score}} & 0.3224 & 0.2136 & 0.1652 & \textbf{0.3464} \\
\hline
\multicolumn{3}{c}{\textbf{Standard Deviation}}& 0.3529 & 0.3444 & 0.2122 & 0.2669 \\
\hline
\end{tabular}
\captionsetup{justification=centering}
\caption{\textcolor{black}{Table showing performance of different financial LLMs on datasets from the FinBen benchmark. All the results reported in the table are without supervised fine-tuning.}}
\label{table:comparison1}
\end{table}

\section{Financial Agent for Data Retrieval and Context Generation}\label{sec:FinancialAgent}
\begin{figure}[h!]
  \centering %{.46\textwidth}
  \includegraphics[width=15 cm]{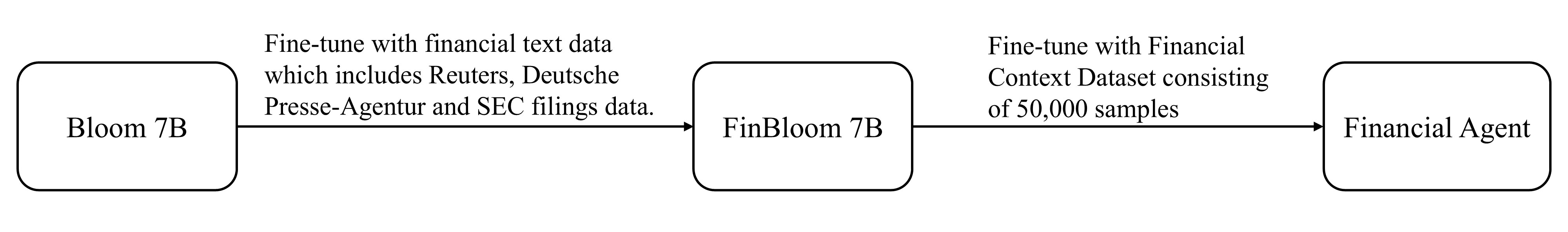}
  \caption{\textcolor{black}{Development process for Financial Agent.}}
  \label{fig:fa-block}
\end{figure}

This section describes the Financial Agent's structure and function, explaining how it works with the Data Module to provide relevant context for an LLM to answer user queries. The Financial Agent was developed by fine-tuning the FinBloom 7B model on the the custom dataset detailed in Section \ref{sec:fincontextdataset}. \textcolor{black}{The entire development process for the Financial Agent has been represented through blocks in Figure~\ref{fig:fa-block}. For comparison purposes, we also fine-tuned the Bloom 7B model directly on Financial Context Dataset and compared it's performance with FinBloom 7B model fine-tuned on the same dataset.}
The fine-tuning process employed the parameter-efficient Prompt Tuning methodology as proposed by \citep{lester2021power}. We split our dataset in a 4 to 1 ratio for training and evaluation sets. We trained the models\textcolor{black} for 5 epochs.  The scores achieved for the training and evaluation sets are reported in Table~\ref{tab:lossPerplexity}. Clearly, the performance of fine-tuned FinBloom 7B is better than fine-tuned Bloom 7B on all metrics.

\begin{table}[h!]
    \centering
    \begin{tabular}{|c|c|c|c|c|}
        % \noline{} & \noline{Loss} & \noline{Perplexity} \\
        \hline
         & \multicolumn{2}{c|}{\textbf{Training Set}} & \multicolumn{2}{c|}{\textbf{Evaluation Set}}\\
        \hline
        \textbf{Model} & \textbf{Loss} & \textbf{Perplexity} & \textbf{Loss} & \textbf{Perplexity}\\
        \hline
            \textcolor{black}{Bloom 7B} & \textcolor{black}{0.0487} & \textcolor{black}{1.0499} & \textcolor{black}{0.0450} & \textcolor{black}{1.0460} \\
        \hline
        FinBloom 7B & 0.0375 & 1.0382 & 0.0076 & 1.0076 \\
        \hline
    \end{tabular}
    \caption{Loss and perplexity scores for training and evaluation data for models fine-tuned on Financial Context Dataset.}
    \label{tab:lossPerplexity}
\end{table}

To further assess the accuracy of the fine-tuned models\textcolor{black} on our custom dataset, we generated an additional 10,000 natural queries and their corresponding labels to create a test set. We then passed the same queries to the models\textcolor{black} as inputs and recorded the outputs. Considering these recorded outputs from the model as the generated text and the original labels in the test set as the reference text, we calculated the similarity between the two sets using the BLEU (Bigram Language Evaluation Understudy) (\cite{papineni2002bleu}) and ROUGE (Recall-Oriented Understudy for Gisting Evaluation) (\cite{lin2004rouge}) metrics. These metrics provide a numerical score reflecting the degree of overlap between the model-generated text and the expected response, which are reported in Table~\ref{tab:bleuRouge}. As we can see from these scores, the fine-tuned FinBloom 7B model is excellent in identifying the required data to answer the queries. \textcolor{black}{The domain-specific FinBloom 7B model significantly outperforms the general Bloom 7B model after both are fine-tuned on Financial Context Dataset, showing the impact of financial domain pre-training.}

\begin{table}[h!]
\begin{center}
 \begin{tabular}{ |c|c|c|c|c| } 
 \hline
 \textbf{Model} & \textbf{BLEU} & \textbf{ROUGE-1 (F1)} & \textbf{ROUGE-2 (F1)} &\textbf{ROUGE-L (F1)} \\
 \hline
 \textcolor{black}{Bloom 7B} & \textcolor{black}{0.8752} & \textcolor{black}{0.8947} & \textcolor{black}{0.8863} & \textcolor{black}{0.8943}\\
 \hline
 FinBloom 7B & 0.9614 & 0.9774 & 0.9693 & 0.9771\\ 
 \hline
\end{tabular}   
\caption{BLUE and ROUGE scores for 10,000 test samples corresponding to Financial Context Dataset.}
\label{tab:bleuRouge}
\end{center}
\end{table}
%Following this, the model outputs (see Column 2 in Table \ref{tab:snippet}) require some reformatting to conform to the data module's required format (see Column 3 in Table \ref{tab:snippet}). The date ranges in the model outputs will remain in the same format as it is in the corresponding query. This was done to allow the model to easily extract the dates/date ranges from the user's query.
%The extracted dates/date ranges were converted to the appropriate format for the structured data requests using a proprietary date parser. The structured data requests can then be passed onto the Data Module. 

\section{Work Flow of Various Modules}\label{sec:workflow}
%Check this, especially the beginning, as now this is a new section xxx.
In this section, we describe in detail the functionality of different modules in our framework and the process of how they interact with each other. Our multi-module system is designed to process and respond to financial queries efficiently and accurately, leveraging a specialized Financial Agent and a robust Data Repository. The core of this system is the seamless communication and collaboration between these modules, ensuring a cohesive and dynamic flow of information from the initial user query to the final response generated by the LLM.

When a user query is received, it is first handled by the Financial Agent that extracts relevant information from the Data Module to handle the user query appropriately.
The Data Module serves as the repository for financial data, encompassing tabular financial information as well as real-time news data pertaining to diverse companies, sectors, and industries. It operates in a dynamic fashion, regularly ingesting and incorporating fresh data to ensure that the information it provides remains current and up-to-date. 
Upon receiving a data request from the Financial Agent, the Data Module constructs a dataframe containing pertinent financial data, conforming to the specifications of the data request. The Data Module provides the relevant news data based on semantic matching of news items with the user query. 
{\color{black}
We use RoBERTa-based \citep{liu2019roberta} semantic search to detect similarity between the requested query and the news headlines.
We develop FinRoBERTa\footnote{FinRoBERTa-A transformer model for finance: \url{https://huggingface.co/Chaitanya14/finroberta}} by adapting the RoBERTa-large transformer model (about 400M parameters) to the financial domain through fine-tuning. RoBERTa-large is fine-tuned for 2 epochs on a mixture of financial text sources, namely Reuters, DPA, and SEC data, to strengthen domain-specific language understanding and also on financial sentiment datasets (\cite{malo2014good,sinha2022sentfin}) %In the second stage, the model is further fine-tuned for 4 epochs on a Headlines dataset to improve robustness for downstream classification settings.
For evaluation, we compare three models: (i) the base RoBERTa-large, (ii) FinBERT, and (iii) our domain-adapted FinRoBERTa. All models are tested on the FOMC dataset (part of FinBen) for three-way stance classification (\textit{hawkish}, \textit{dovish}, \textit{neutral}). Table~\ref{tab:fomc_results} reports the results, demonstrating that FinRoBERTa substantially outperforms both FinBERT and the vanilla RoBERTa-large baseline in terms of F1 score and accuracy. It is noteworthy that FinRoBERTa and FinBERT have been fine-tuned on financial sentiments but RoBERTA-large is not fine-tuned on any sentiment data. However, none of these models have been exposed to the FOMC dataset (and other datasets in FinBen) during the training/fine-tuning phases.
\begin{table}[t]
\centering
\caption{{\color{black}Performance comparison on FOMC stance classification (hawkish/dovish/neutral).}}
\label{tab:fomc_results}
\begin{tabular}{lcc}
\hline
\textbf{Model} & \textbf{F1 Score} & \textbf{Accuracy} \\
\hline
FinRoBERTa (RoBERTa-large, domain-adapted) & 0.7674 & 0.7661 \\
FinBERT & 0.5098 & 0.4960 \\
RoBERTa-large & 0.4258 & 0.4136 \\
\hline
\end{tabular}
\end{table}
}%\color ends
%The data module contains the financial metrics, based on their reporting frequency, and leverages this knowledge to effectively present data. 
The Data Module also provides further details, like company/industry specific financial details for the requested date ranges. 
%, the module identifies the closest date range(s) within which the financial data is available. 
For example, net income is reported quarterly, so the module presents the quarterly net incomes for the closest available date range to the one in the request. Share price, on the other hand, is reported at a higher frequency, so the module presents the share price data for the exact date range as requested in the query. If there are no dates/date ranges present in the request, then the latest date range for which the data is available is chosen. 
%The module then retrieves the relevant metric data for that date range and in addition, communicates the corresponding date range and frequency for which this metric's data is provided. 
Finally, the agent integrates this data with the initial user query and supplies the enriched query to the LLM for processing. 
%the LLM receives accurate information regarding the metric and its associated date range. 
%As mentioned earlier, for retrieval of the news data, we use RoBERTa-based semantic search. 
%An illustrative instance of a data request alongside the corresponding response generated by the module can be seen in the example presented in Appendix \ref{appendix:example}.

\begin{figure}[h!]
  \centering %{.46\textwidth}
  \includegraphics[width=16 cm]{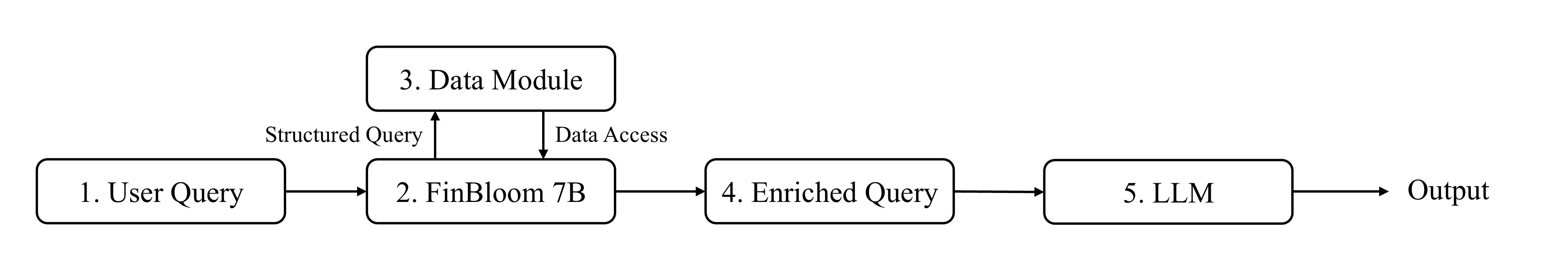}
  \caption{\textcolor{black}{Work flow and interoperability of the modules.}}
  \label{fig:flow-block}
\end{figure}

The data flow and the interoperability of the modules are shown through Figure~\ref{fig:flow-block} and provided in steps below.
\begin{enumerate}
    \item The user's initial query, $x$, is received by the Financial Agent.
    \item The Financial Agent identifies the required data to answer this query and converts it into a structured data request to be passed onto the Data Module.
    \item The Data Module receives this data request and returns the relevant financial and news data.
    \item This data is converted into a string form \textit{c} and appended to the initial query \textit{x} to come up with an enriched query. The enriched query is sent to the LLM to answer the given financial query using the data provided.
    \item The LLM receives the contextualized and enriched query, $(c,x)$, to generate the output.
\end{enumerate}
An illustrative example to demonstrate how the Financial Agent architecture works in knowledge-grounding of LLMs and helps to answer financial queries comprehensively has been provided in Example 1 in Appendix \ref{appendix:example}. \textcolor{black}{In summary, we chose the architecture shown in Figure~\ref{fig:flow-block} over alternatives for several key reasons. Although fine-tuning a general-purpose LLM like GPT-3.5 with financial data is possible, we opted for a domain-specific model to ensure greater control over training, customization, and deployment costs--crucial for real-time financial applications--and to enable faster inference. We also deliberately kept the Data Module separate from the Financial Agent to allow for a modular system design, ensuring flexibility in data updates without needing to retrain the agent frequently. Recent updates in general-purpose LLM allow web access. However, it is inadequate for high-volume, high-velocity financial contexts. These models often lack access to structured financial data (like intraday trading data, earnings call transcripts, real-time market feeds, regulatory filings, etc.) and are prone to hallucinations or approximations when precise figures are needed (like revenue growth, dividend yield, volume spikes, sectoral indices, etc.). These issues of hallucinations and approximations are demonstrated in empirical evaluations illustrated in Examples 2 and 3 in Appendix~\ref{appendix:example}. Finally, Example 4 in Appendix~\ref{appendix:example} compares the performance of an LLM with web access against the same LLM operating in our proposed architecture.}

\section{Conclusions}\label{sec:conclusions}
%This study highlights the importance of knowledge-grounding in LLMs, particularly for interactive tasks that demand access to real-time information. The paper makes multiple contributions to efficiently achieve knowledge-grounding in finance. We developed a Financial Context Dataset comprising over 50,000 financial queries paired with the contextual information necessary for their resolution. We then trained FinBloom 7B, a 7 billion-parameter language model, on 14 million financial news articles sourced from Reuters and Deutsche Presse-Agentur (DPA) and 12 million documents sourced from SEC filings. Finally, we fine-tuned FinBloom 7B on the Financial Context Dataset, creating a Financial Agent capable of generating the required context for user queries and retrieving it from a data module to handle financial queries efficiently. Multiple interactions with large-scale LLMs (for example, GPT 3.5 or GPT 4) may lead to high latency for financial decision making. To avoid this, a Financial Agent has been trained that can be paired with any large-scale LLM. The user provides the query, the Financial Agent understands the context and retrieves data, and finally the large-scale LLM processes the contextual information and query. The framework has low latency and therefore better capability in handling high velocity data. 

This study emphasizes the role of knowledge-grounding in LLMs, particularly for interactive tasks that require real-time information access. 
%It presents multiple contributions toward efficient knowledge-grounding in the domain of finance.
First, we developed the Financial Context Dataset, which comprises over 50,000 financial queries paired with the contextual information necessary for their resolution. Building on this foundation, we trained FinBloom 7B, a 7 billion parameter language model, using a corpus of over 14 million financial news articles sourced from Reuters and Deutsche Presse-Agentur (DPA) and 12 million documents from SEC filings. Subsequently, we fine-tuned FinBloom 7B on the Financial Context Dataset to create a Financial Agent. This agent is capable of generating the required context for user queries and efficiently retrieving it from a data module to address financial queries.
A key challenge in financial decision-making lies in mitigating the high latency associated with multiple interactions with large-scale LLMs (e.g., GPT-3.5 or GPT-4). To address this, the fine-tuned Financial Agent seamlessly integrates with any large-scale LLM. In our proposed framework, the user submits a query, the Financial Agent processes and retrieves the necessary context, and the LLM subsequently analyzes the query along with the contextual information. This design ensures low latency, enabling effective handling of high-velocity financial data effectively.

Future research could expand this framework into a multi-agent system, integrating additional agents such as financial video analysis tools to further enhance its capabilities. With these advancements, our work holds the potential to transform the delivery of financial services, making them more efficient, personalized, and accessible to a broader audience.

\textcolor{black}{
\section{Acknowledgements}
The authors gratefully acknowledge the valuable interactions and contributions of Mr. Sougata Basu (CEO, CashRich) and Mr. Nayan Agarwal (Founder, Beehive Wealth) towards this study. The authors also acknowledge Krishnamurthy Tandon School of AI and Brij Disa Centre for Data Science and AI at Indian Institute of Management Ahmedabad for supporting this study.
}

\vskip 0.2in
\bibliography{references}

\newpage
\renewcommand{\theHsection}{A\arabic{section}}
\appendix
{\color{black}
\section{Data Module Implementation and Retrieval Details}\label{app:data-module}
The Data Module is implemented as a logical abstraction layer that unifies access to two physically distinct backends optimized for different access patterns: (i) a relational store for structured/tabular financial information and (ii) a document store with a vector index for news. The Financial Agent interacts with the Data Module through a single, structured request interface (cf.\ the ``Structured Data Request'' format in Table~\ref{tab:snippet}). Internally, the Data Module routes sub-requests to the appropriate backend, then merges the results into (a) a tabular result (as a \texttt{pandas} DataFrame) and (b) a ranked list of relevant news items. These outputs are subsequently linearized into the textual context $c$ used by the language model (as illustrated in Figure~\ref{fig:Financial Agent-withkg}).

\subsection{Tabular Data Store}
Structured financial data (e.g., fundamental metrics, prices, and other numeric series) are stored in a relational database (PostgreSQL). The schema follows a star design to support efficient filtering by company, metric, and date range:
\begin{itemize}
    \item \texttt{companies(company\_id, ticker, name, sector, \dots)}
    \item \texttt{metrics(metric\_id, name, unit, frequency, \dots)}
    \item \texttt{metric\_values(company\_id, metric\_id, date, value, updated\_at, \dots)}
\end{itemize}
This organization supports the dominant query pattern induced by the structured request representation: retrieving specific metrics for specific companies over a specified date interval.

\subsection{News Data Store}
News are stored as documents in a single Elasticsearch index that serves as both a document store and a vector-search backend. Each document contains structured metadata alongside a dense embedding:
\begin{itemize}
    \item \texttt{article\_id} (ID), \texttt{source} (provider), \texttt{published\_at} (timestamp)
    \item \texttt{headline}, \texttt{body}
    \item \texttt{tickers} and \texttt{sectors} (entity tags for filtering)
    \item \texttt{embedding}: a 768-dimensional dense vector field (stored as \texttt{dense\_vector}).
\end{itemize}
Embeddings are pre-computed using a RoBERTa-based encoder \cite{liu2019roberta} fine-tuned on financial news, and stored directly in the index to enable ANN retrieval with cosine similarity.

\subsection{Update Cadence by Data Type}
The Data Module supports different update cadences depending on the data source:
\begin{itemize}
    \item \textbf{Market data (prices, volumes).} Intraday prices and volumes are ingested in micro-batches from the data provider; in our deployment this results in an approximate lag of \textbf{30 minutes} from exchange publication to availability within the Data Module.
    \item \textbf{Fundamental metrics and statements.} Quarterly and annual fundamentals are ingested via \textbf{daily batch jobs}, which is sufficient for the decision-support scenarios targeted in this work.
    \item \textbf{News articles.} News feeds are polled at short intervals (\textbf{every 30 minutes}); newly arrived items are tokenized, embedded, and indexed for vector search. It is ensured that duplicate news items do not get stored.
\end{itemize}
This describes a research-grade implementation of the logical data layer. An industry deployment can reduce latency further by adopting higher-frequency ingestion and streaming infrastructure.

%\subsection{Historical Queries and Data Versioning}
For structured metrics, the relational store is treated as append-only, i.e. updates do not overwrite existing rows. Instead, new records are inserted with an \texttt{updated\_at} timestamp. When answering historical queries (i.e. queries specifying an explicit past interval, for instance, `from Apr 2016 to Jul 2017' in Query 3, Table~\ref{tab:snippet}), the Data Module filters as per the requested date range or the recent values in case date range is not specified. For news, the system preserves the full history of articles. Retrieval first applies filters on \texttt{published\_at} and \texttt{tickers}/\texttt{sectors}, and only then applies semantic similarity ranking.

%\subsection{From Structured Requests to Tabular SQL Queries}
The ``Structured Data Request'' representation encodes: (i) companies, (ii) metrics (and related metrics), (iii) a date range. When the Financial Agent processes a user query, it generates the same structured representation. The Data Module then executes the following steps:
\begin{enumerate}
    \item \textbf{Normalize company entities.} Company strings are matched to internal \texttt{company\_id}s using canonical mappings (ticker$\rightarrow$company) together with a curated alias list (alternative names and abbreviations).
    \item \textbf{Normalize metrics.} Metric strings are mapped to \texttt{metric\_id}s in a metrics catalog that encodes frequency (daily/quarterly/annual) and the target table. If the request includes related metrics, these are expanded into additional \texttt{metric\_id}s.
    \item \textbf{Parse dates into explicit intervals.} Date ranges are converted into explicit start--end intervals, enabling consistent filtering of time series. If date range is not provided in the query, it is assumed that the query is related to the most recent dates.
    \item \textbf{Issue parameterized SQL.} For each \texttt{(company\_id, metric\_id)} pair, the Data Module issues parameterized SQL restricted to the requested interval. For quarterly metrics, values are aligned to the nearest reporting period as described in Section~\ref{sec:workflow}.
\end{enumerate}

\subsection{News Retrieval via RoBERTa-Based Semantic Search (FinRoBERTa)}
\label{app:retrieval-news}
News retrieval follows a two-stage pipeline:
\begin{enumerate}
    \item \textbf{Embedding model.} A RoBERTa-based encoder \cite{liu2019roberta} is fine-tuned on financial news headlines (we refer to this as FinRoBERTa) and short snippets using a contrastive loss objective. The model produces 768-dimensional embeddings for both queries and headlines.
    \item \textbf{Filtered vector search.} All headlines are embedded offline and stored in Elasticsearch as a \texttt{dense\_vector} field. At query time, the query is embedded and used to retrieve the top-$k$ nearest neighbours by cosine similarity using approximate kNN search, subject to filters (ticker, sector and date range constraints) derived from the structured request.
\end{enumerate}

%\paragraph{Thresholding and context control.}
We tune the retrieval hyperparameters ($k=20$ and a cosine similarity threshold $t=0.2$) on a held-out validation set of (query, relevant-article) pairs. Articles with similarity below $t$ are discarded to avoid introducing weakly related news into the context.  We keep the number of headlines per query small to control the context length.
%In practice, we include at most 20 headlines per query to control context length.

%\subsection{Missing and Ambiguous Data}
%If the requested tabular data are missing (e.g., metric not available for a company, or no coverage in the requested date interval), the Data Module returns explicit missing indicators (e.g., NULL/empty results) so the Financial Agent can (i) answer using the available subset, (ii) fall back to closely related metrics when appropriate, or (iii) state that the requested data are unavailable for the specified scope.

%For ambiguous company/metric strings, the Data Module prioritizes exact ticker matches when available. Otherwise, it uses the curated alias list and metrics catalog constraints to disambiguate; if ambiguity remains unresolved, the module returns multiple candidates (or a structured ambiguity signal) rather than injecting a potentially incorrect value.

%\subsection{Retrieval-Level Evaluation}
%In addition to downstream task evaluation, we include a retrieval-level evaluation that measures the quality of (i) tabular retrieval and (ii) news retrieval under the same filters and thresholds used during inference. Specifically, we evaluate retrieval precision and recall at $k$ on a validation set of (query, relevant-item) pairs to quantify how often the Data Module returns the intended supporting evidence and to guide tuning of $k$ and the similarity threshold $t$.
}%Color green ends

\newpage

\section{Query Collection, Templatization and Sampling}\label{sec:templateApp}
This section details the methodology and key insights gained from interviews with financial advisors and retail investors. It outlines the diverse backgrounds of the interviewees and how their experiences and priorities informed the structure and content of our comprehensive Financial Context Dataset.
 \textcolor{black}{Our interviews with financial advisors and retail investors provided valuable insights that were instrumental in constructing a comprehensive dataset of financial queries. The advisors we spoke with represented a diverse professional landscape, including those working for the wealth management divisions of banks serving their established customer base, and independent private advisors catering to individual clients. There were 4 advisors working for banks and 6 were independent advisors. Interestingly, most of the independent advisors had transitioned to their current practice from a professional setup, where they gained experience as wealth managers typically for more than 5 years. The educational backgrounds of all advisors were robust, encompassing Masters in Business Administration (MBA), Masters in Finance, Chartered Financial Analysts (CFA) and Chartered Accountancy (CA), with some even holding engineering and science degrees, highlighting a strong analytical foundation. All the advisors interviewed managed more than 100 relationships. All of the advisors that we interviewed were registered with the regulator (Securities and Exchange Board of India) after passing exams specified by the regulatory authority. The eligibility requires individuals to have Bachelor's/Master's degrees or professional qualifications. The role of advisors, while overlapping in certain functions, differs significantly from that of another category of individuals-relationship managers-who are often unregistered with the regulatory authority.}
%A common thread was the significant experience required to establish a practice, with most advisors needing 4-5 years in the field before venturing out on their own.
\textcolor{black}{These interviews shed light on several critical aspects of a financial advisor's practice, directly informing the types of queries included in our dataset. A primary focus for advisors when engaging with new clients is to first ascertain if the client has made investments in the past, to what extent they understand investments, what is their risk profile, and what are their goals. This is followed by understanding their liquidity requirements, investment horizons, preferences for asset classes and overall financial situation to tailor a suitable portfolio. The client base itself is varied, ranging from ``passive investors'' who delegate most of the investment decisions to advisors due to a lack of technical market knowledge or time, to ``active investors'' who engage in frequent and in-depth technical discussions about company financials and play a more hands-on role in decision-making. This distinction is crucial for understanding the varying levels of detail and technicality required in responses to client queries. Furthermore, the advisors consistently emphasized the paramount importance of staying abreast of the latest financial news. They highlighted that any significant event, whether micro or macroeconomic, could have a direct and immediate impact on their clients' portfolios, necessitating constant monitoring, analysis and update. This underscored the need for the dataset to include queries that require real-time financial data and event-driven information to provide comprehensive and timely answers. The retail investors whom we interviewed primarily represented three classes: 10\% large investments (more than USD 250,000 investments), 20\% medium investments (between USD 50,000 to 250,000 investments), 70\% low investments (less than USD 50,000 investments). All investments were in Indian Rupees (INR) that we have converted approximately into US dollars (USD). There was variability in risk profile of the bottom 70\% investors in terms of home ownership, family size, and income.} 
%A few lines to maintain continuity XXX

Following this initial collection, 5,000 queries were manually annotated with contextual information. These annotated queries were then templatized and expanded to a dataset of 50,000 using a systematic sampling methodology. Table~\ref{tab:template2} provides templates and the corresponding queries generated from it. The table shows how 5000 templates have been used to synthesize (sample) a dataset with 50000 queries.
\begin{table}[t]
\footnotesize
\centering
\begin{tabular}{|p{0.25\textwidth}|p{0.65\textwidth}|}
\hline
\textbf{Template} & \textbf{Generated Queries} \\ \hline
\textit{Judging by [company]'s [metrics][date], is it a good investment opportunity?} & 
\begin{itemize}
    \item Judging by Tesla's market capitalization growth in 2023, is it a good investment opportunity?
    \item Judging by Google's revenue in Q1 2024, is it a good investment opportunity?
\end{itemize} \\ \hline
\textit{Evaluate the [metrics] associated with [company][date].} & 
\begin{itemize}
    \item Evaluate the customer retention rates associated with Netflix in 2023.
    \item Evaluate the price-to-earnings ratio associated with TCS in 2022.
\end{itemize} \\ \hline
\textit{List the top 5 companies [date] in terms of [metrics].} & 
\begin{itemize}
    \item List the top 5 companies in 2024 in terms of annual revenue.
    \item List the top 5 companies in Q3 2023 in terms of stock price growth.
\end{itemize} \\ \hline
\textit{Were there any improvements seen in the [metrics] of [company1][date] as compared to [company2]?} & 
\begin{itemize}
    \item Were there any improvements seen in the renewable energy adoption metrics of Shell in 2023 as compared to ExxonMobil?
    \item Were there any improvements seen in the employee satisfaction scores of Microsoft in 2023 as compared to Google?
\end{itemize} \\ \hline
\end{tabular}
\caption{Templates and Realistic Queries}
\label{tab:template2}
\end{table}

\section{FinBen Benchmark and Comparison Results}\label{sec:finben}
To rigorously evaluate our model's performance across a wide spectrum of financial tasks, we employed the comprehensive FinBen (\cite{xie2024finben}) benchmark. This benchmark comprises a collection of datasets categorized by the specific large language model (LLM) abilities they are designed to evaluate. These categories are as follows:
\begin{enumerate}
\item Information Extraction: This category includes the NER (\cite{alvarado2015domain}), FINER-ORD (\cite{shah2023finer}), FinRED (\cite{sharma2022finred}), SC(\cite{mariko2020financial}), CD (\cite{mariko2020financial}), FNXL (\cite{sharma2023financial}), and FSRL (\cite{lamm2018textual}) datasets, designed to assess the LLM's capacity for extracting structured information from unstructured text.
\item Textual Analysis: This section focuses on evaluating textual understanding and analytical capabilities, utilizing the Financial Phrase Bank (FPB) (\cite{malo2014good}), FiQA-SA (\cite{maia2018financial}), TSA (\cite{cortis-etal-2017-semeval}), Headlines (\cite{sinha2021impact}), FOMC (\cite{shah2023trillion}), FinArg-AUC (\cite{sy2023fine}), FinArg-ARC (\cite{sy2023fine}), Multifin (\cite{jorgensen2023multifin}), MA (\cite{yang2020generating}), and MLESG (\cite{chen2023multi}) datasets.
\item Question Answering: This category assesses the LLM's ability to comprehend and answer questions based on given text, employing the FinQA (\cite{chen2021finqa}), TATQA (\cite{zhu2021tat}), Regulations (\cite{xie2024finben}), and ConvFinQA (\cite{chen2022convfinqa}) datasets.
\item Text Generation: This section evaluates the LLM's capacity to generate coherent and contextually relevant text, utilizing the EDTSUM (\cite{xie2024finben}) and ECTSUM (\cite{mukherjee2022ectsum}) datasets.
\item Forecasting: This category tests the LLM's ability to predict future trends or outcomes, using the BigData22 (\cite{soun2022accurate}), ACL18 (\cite{xu2018stock}), and CIKM18 (\cite{wu2018hybrid}) datasets.
\item Risk Management: This section evaluates the LLM's ability to assess and manage risk, leveraging datasets such as German (\cite{statlog_(german_credit_data)_144}), Australian (\cite{statlog_(australian_credit_approval)_143}), LendingClub(\cite{feng2023empowering}), ccf (\cite{feng2023empowering}), ccfraud (\cite{feng2023empowering}), Polish (\cite{feng2023empowering}), Taiwan (\cite{feng2023empowering}), portoseguro (\cite{feng2023empowering}), and travelinsurance (\cite{feng2023empowering}).
\end{enumerate}

Table~\ref{tab:allComparison} provides an extensive comparison of multiple LLMs on FinBen benchmark that consists of 35 datasets. \textcolor{black}{Please note that the MCC metric can be negative, as it ranges from -1 to +1. The evaluation results were extracted and analyzed using a standardized procedure. The maximum number of generation tokens was set to 1024, consistent with the approach followed in the FinBen paper. For datasets requiring specific output formats, such as Named Entity Recognition (NER) or classification tasks, regular expressions were employed to extract outputs in the required structure; if extraction was not feasible, the response was treated as null. In datasets demanding a single word or numerical answer, the text preceding the first whitespace character in the generated output was used. For long-form question answering tasks, the entire generated text was taken as the final response. Each model was run three times, with equal weightage assigned to each run, and the final scores were computed as the average across the runs. All models were evaluated using greedy decoding, and decoding settings remained unchanged throughout the evaluation process.
}
\begin{table}
\centering
\setlength{\arrayrulewidth}{0.1mm}
\setlength{\tabcolsep}{2pt}
\renewcommand{\arraystretch}{1.5}
\fontsize{5pt}{5pt}\selectfont
\rotatebox{90}{
\begin{tabular} 
{ m{1.6cm} m{3.3cm} m{1.0cm} m{1.1cm} m{0.8cm} m{0.9cm} m{1.1cm} m{1.0cm} m{1.0cm} m{0.9cm} m{1.0cm} m{1.1cm} m{0.7cm} m{0.9cm} m{1.3cm} m{1.1cm} }
\hline
\textbf{Dataset} & Task & \textbf{Metrics} & \textbf{ChatGPT} & \textbf{GPT 4} & \textbf{Gemini} & \textbf{LLaMA2 7B-Chat} & \textbf{LLaMA2 70B} & \textbf{LLaMA3 8B} & \textbf{FinMA 7B} & \textbf{FinGPT 7B-lora} & \textbf{InternLM 7B} & \textbf{Falcon 7B} & \textbf{Mixtral 7B} & \textbf{CFGPT \newline sft-7B-Full} & \textbf{FinBloom 7B} \\
\hline
NER & named entity recognition & EntityF1 & 0.77 & \textbf{0.83} & 0.61 & 0.18 & 0.04 & 0.08 & 0.69 & 0.00 & 0.00 & 0.00 & 0.24 & 0.00 & 0.00 \\
FINER-ORD & named entity recognition & EntityF1 & 0.28 & \textbf{0.77} & 0.14 & 0.02 & 0.07 & 0.00 & 0.00 & 0.00 & 0.00 & 0.00 & 0.05 & 0.00 & 0.00 \\
FinRED & relation extraction & F1 & 0.00 & \textbf{0.02} & 0.00 & 0.00 & 0.00 & 0.00 & 0.00 & 0.00 & 0.00 & 0.00 & 0.00 & 0.00 & 0.00 \\
SC & causal classification & F1 & 0.80 & 0.81 & 0.74 & 0.85 & 0.61 & 0.69 & 0.19 & 0.00 & \textbf{0.88} & 0.67 & 0.83 & 0.15 & 0.84 \\
CD & causal detection & F1 & 0.00 & 0.01 & \textbf{0.03} & 0.00 & 0.01 & 0.00 & 0.00 & 0.00 & 0.00 & 0.00 & 0.00 & 0.00 & 0.00 \\
FNXL & numeric labeling & EntityF1 & 0.00 & 0.00 & 0.00 & 0.00 & 0.00 & 0.00 & 0.00 & 0.00 & 0.00 & 0.00 & 0.00 & 0.00 & 0.00 \\
FSRL & textual analogy parsing & EntityF1 & 0.00 & 0.01 & \textbf{0.03} & 0.00 & 0.01 & 0.00 & 0.00 & 0.00 & 0.00 & 0.00 & 0.00 & 0.00 & 0.00 \\
\hline
\multirow{2}{*}{FPB}  & \multirow{2}{*}{sentiment analysis} & F1 & 0.78 & 0.78 & 0.77 & 0.39 & 0.73 & 0.52 & \textbf{0.88} & 0.00 & 0.69 & 0.07 & 0.29 & 0.35 & 0.32 \\
                      &  &  Acc & 0.78 & 0.76 & 0.77 & 0.41 & 0.72 & 0.52 & \textbf{0.88} & 0.00 & 0.69 & 0.05 & 0.37 & 0.26 & 0.31 \\
FiQA-SA & sentiment analysis & F1 & 0.60 & 0.80 & 0.81 & 0.76 & \textbf{0.83} & 0.70 & 0.79 & 0.00 & 0.81 & 0.77 & 0.16 & 0.42 & 0.47 \\
TSA & sentiment analysis & RMSE & 0.53 & 0.50 & 0.37 & 0.71 & 0.57 & 0.25 & 0.80 & 0.00 & \textbf{0.29} & 0.50 & 0.50 & 1.05 & 0.77 \\
Headlines & news headline classification & AvgF1 & 0.77 & 086 & 0.78 & 0.72 & 0.63 & 0.60 & \textbf{0.97} & 0.60 & 0.60 & 0.45 & 0.60 & 0.61 & 0.45 \\
\multirow{2}{*}{FOMC}  & \multirow{2}{*}{hawkish-dovish classification} & F1 & 0.64 & \textbf{0.71} & 0.40 & 0.35 & 0.49 & 0.40 & 0.49 & 0.00 & 0.36 & 0.30 & 0.37 & 0.16 & 0.20\\
                      &  & Acc & 0.60 & \textbf{0.69} & 0.60 & 0.49 & 0.47 & 0.41 & 0.46 & 0.00 & 0.35 & 0.30 & 0.35 & 0.21 & 0.25 \\

FinArg-AUC & argument unit classification & MicroF1 & 0.50 & 0.60 & 0.31 & 0.46 & 0.58 & 0.51 & 0.27 & 0.00 & 0.39 & 0.23 & 0.39 & 0.05 & \textbf{0.81} \\
FinArg-ARC & argument relation classification & MicroF1 & 0.39 & 0.40 & \textbf{0.60} & 0.27 & 0.36 & 0.28 & 0.08 & 0.00 & 0.33 & 0.32 & 0.57 & 0.05 & 0.20 \\
MultiFin & multi-class classification & MicroF1 & 0.59 & \textbf{0.65} & 0.62 & 0.20 & 0.63 & 0.39 & 0.14 & 0.00 & 0.34 & 0.09 & 0.37 & 0.05 & 0.25 \\
MA & deal completeness classification & MicroF1 & 0.85 & 0.79 & 0.84 & 0.70 & \textbf{0.86} & 0.34 & 0.45 & 0.00 & 0.78 & 0.39 & 0.34 & 0.25 & 0.48 \\
MLESG & ESG Issue Identification & MicroF1 & 0.25 & \textbf{0.35} & 0.34 & 0.03 & 0.31 & 0.12 & 0.00 & 0.00 & 0.14 & 0.06 & 0.17 & 0.01 & 0.07 \\
\hline
FinQA & question answering & EmAcc & 0.58 & \textbf{0.63} & 0.00 & 0.00 & 0.06 & 0.00 & 0.04 & 0.00 & 0.00 & 0.00 & 0.00 & 0.00 & 0.00 \\
TATQA & question answering & EmAcc & 0.00 & 0.13 & \textbf{0.18} & 0.03 & 0.01 & 0.01 & 0.00 & 0.00 & 0.00 & 0.00 & 0.01 & 0.00 & 0.00 \\
\multirow{2}{*}{Regulations}  & \multirow{2}{*}{long-form question answering} & Rouge-1 & 0.12 & 0.11 & - & 0.24 & - & 0.10 & 0.12 & 0.01 & 0.04 & 0.03 & - & 0.14 & 0.21 \\
                      &  & BertScore & 0.64 & 0.62 & - & 0.65 & - & 0.60 & 0.59 & 0.40 & 0.57 & 0.14 & - & 0.57 & 0.80 \\
ConvFinQA & multi-turn question answering & EmAcc & 0.60 & \textbf{0.76} & 0.43 & 0.00 & 0.25 & 0.00 & 0.20 & 0.00 & 0.00 & 0.00 & 0.31 & 0.01 & 0.00 \\
\hline
\multirow{2}{*}{EDTSUM}  & \multirow{2}{*}{text summarization} & Rouge-1 & 0.17 & 0.20 & \textbf{0.39} & 0.17 & 0.25 & 0.14 & 0.13 & 0.00 & 0.13 & 0.15 & 0.12 & 0.01 & 0.08 \\
                      &  & BertScore & 0.66 & 0.67 & 0.72 & 0.62 & 0.68 & 0.60 & 0.38 & 0.52 & 0.48 & 0.57 & 0.61 & 0.51 & \textbf{0.79} \\
\multirow{2}{*}{ECTSUM}  & \multirow{2}{*}{text summarization} & Rouge-1 & 0.00 & 0.00 & 0.00 & 0.00 & 0.00 & 0.00 & 0.00 & 0.00 & 0.00 & 0.00 & 0.00 & 0.00 & 0.00 \\
                      &  & BertScore & 0.00 & 0.00 & 0.00 & 0.00 & 0.00 & 0.00 & 0.00 & 0.00 & 0.00 & 0.00 & 0.00 & 0.00 & 0.00 \\
\hline
\multirow{2}{*}{BigData22}  & \multirow{2}{*}{stock movement prediction} & Acc & 0.53 & 0.54 & 0.55 & 0.54 & 0.47 & 0.55 & 0.51 & 0.45 & \textbf{0.56} & 0.55 & 0.46 & 0.45 & 0.37 \\
                      &  & MCC & -0.025 & 0.03 & 0.04 & 0.05 & 0.00 & 0.02 & 0.02 & 0.00 & \textbf{0.08} & 0.00 & 0.02 & 0.03 & -0.02 \\
\multirow{2}{*}{ACL18}  & \multirow{2}{*}{stock movement prediction} & Acc & 0.50 & \textbf{0.52} & 0.52 & 0.51 & 0.51 & 0.52 & 0.51 & 0.49 & 0.51 & 0.51 & 0.49 & 0.48 & 0.29 \\
                      &  & MCC & 0.005 & 0.02 & \textbf{0.04} & 0.01 & 0.01 & 0.02 & 0.03 & 0.00 & 0.02 & 0.00 & 0.00 & -0.03 & 0.01 \\
\multirow{2}{*}{CIKM18}  & \multirow{2}{*}{stock movement prediction} & Acc & 0.55 & \textbf{0.57} & 0.54 & 0.55 & 0.49 & 0.57 & 0.50 & 0.42 & 0.57 & 0.47 & 0.42 & 0.41 & 0.44 \\
                      &  & MCC & 0.01 & 0.02 & 0.02 & -0.03 & -0.07 & 0.03 & \textbf{0.08} & 0.00 & -0.03 & -0.06 & -0.05 & -0.07 & -0.01 \\
\hline
\multirow{2}{*}{German}  & \multirow{2}{*}{credit scoring} & F1 & 0.20 & 0.55 & 0.52 & \textbf{0.57} & 0.17 & 0.56 & 0.17 & 0.52 & 0.41 & 0.23 & 0.53 & 0.53 & 0.53 \\
                      &  & MCC & -0.10 & -0.02 & 0.00 & \textbf{0.03} & 0.00 & 0.05 & 0.00 & 0.00 & -0.30 & -0.07 & 0.00 & 0.00 & -0.04 \\
\multirow{2}{*}{Australian}  & \multirow{2}{*}{credit scoring} & F1 & 0.41 & \textbf{0.74} & 0.26 & 0.26 & 0.41 & 0.26 & 0.41 & 0.38 & 0.34 & 0.26 & 0.26 & 0.29 & 0.51 \\
                      &  & MCC & 0.00 & \textbf{0.47} & 0.00 & 0.00 & 0.00 & 0.00 & 0.00 & 0.11 & 0.13 & 0.00 & 0.00 & -0.10 & 0.08 \\
\multirow{2}{*}{LendingClub}  & \multirow{2}{*}{credit scoring} & F1 & 0.20 & 0.55 & 0.65 & \textbf{0.72} & 0.17 & 0.10 & 0.61 & 0.00 & 0.59 & 0.02 & 0.61 & 0.05 & 0.53 \\
                      &  & MCC & -0.10 & -0.02 & \textbf{0.19} & 0.00 & 0.00 & -0.15 & 0.00 & 0.00 & 0.15 & -0.01 & 0.08 & 0.01 & -0.02 \\
\multirow{2}{*}{ccf}  & \multirow{2}{*}{fraud detection} & F1 & 0.20 & 0.55 & 0.96 & 0.00 & 0.17 & 0.01 & 0.00 & 1.00 & \textbf{1.00} & 0.10 & 0.00 & 0.00 & 0.70 \\
                      &   &MCC & -0.10 & -0.02 & -0.01 & \textbf{0.00} & 0.00 & 0.00 & 0.00 & 0.00 & 0.00 & 0.00 & 0.00 & 0.00 & -0.01 \\
\multirow{2}{*}{ccfraud}  & \multirow{2}{*}{fraud detection} & F1 & 0.20 & 0.55 & \textbf{0.90} & 0.25 & 0.17 & 0.36 & 0.01 & 0.00 & 0.57 & 0.62 & 0.48 & 0.03 & 0.77 \\
                      &  & MCC & -0.10 & -0.02 & 0.00 & -0.16 & 0.00 & -0.03 & -0.06 & 0.00 & -0.13 & -0.02 & \textbf{0.16} & 0.01 & -0.01 \\
\multirow{2}{*}{polish}  & \multirow{2}{*}{financial distress identification} & F1 & 0.20 & 0.55 & 0.86 & 0.92 & 0.17 & 0.83 & 0.92 & 0.30 & 0.92 & 0.76 & \textbf{0.92} & 0.40 & 0.30 \\
                      &  & MCC & -0.10 & -0.02 & \textbf{0.14} & 0.00 & 0.00 & -0.06 & -0.01 & 0.00 & 0.07 & 0.05 & 0.00 & -0.02 & 0.00 \\
\multirow{2}{*}{taiwan}  & \multirow{2}{*}{financial distress identification} & F1 & 0.20 & 0.55 & \textbf{0.95} & 0.95 & 0.17 & 0.26 & 0.95 & 0.60 & 0.95 & 0.00 & 0.95 & 0.70 & 0.32 \\
                      &  & MCC & -0.10 & -0.02 & \textbf{0.00} & -0.01 & 0.00 & -0.07 & 0.00 & -0.02 & -0.01 & 0.00 & 0.00 & 0.00 & 0.02 \\
\multirow{2}{*}{portoseguro}  & \multirow{2}{*}{claim analysis} & F1 & 0.20 & 0.55 & 0.95 & 0.01 & 0.17 & 0.94 & 0.04 & \textbf{0.96} & 0.96 & 0.95 & 0.72 & 0.00 & 0.41 \\
                      &  & MCC & -0.10 & -0.02 & 0.00 & -0.05 & 0.00 & -0.01 & \textbf{0.01} & 0.00 & 0.00 & 0.00 & 0.01 & 0.00 & 0.00 \\
\multirow{2}{*}{travelinsurance}  & \multirow{2}{*}{claim analysis} & F1 & 0.20 & 0.55 & 0.00 & 0.00 & 0.17 & 0.00 & 0.00 & \textbf{0.98} & 0.89 & 0.77 & 0.00 & 0.03 & 0.50 \\
                      &  & MCC & -0.10 & -0.02 & 0.00 & 0.00 & 0.00 & 0.00 & 0.00 & 0.00 & \textbf{0.12} & -0.03 & 0.00 & 0.01 & 0.01 \\
\hline
\end{tabular}}
\captionsetup{justification=centering}
\caption{Performance of LLMs on the FinBen benchmark as average of three runs.}
\label{tab:allComparison}
\end{table}
\newpage
\section{Knowledge-Grounding an LLM using Financial Agent}\label{appendix:example}
% the \\ insures the section title is centered below the phrase: AppendixA
We present multiple examples in this section where financial queries are posed to LLMs with different settings. To begin with, we  present an illustrative example showcasing the Financial Agent Framework processing a user's query, building the relevant context, constructing the modified query and then interacting with an LLM (ChatGPT), which does not interact with the web. The query was posed in September 2023.
%GPT-3.5 has been used for producing the final  generated by GPT-3.5 to the modified query:\\\\
\\ \\
\noindent\textbf{Example Query 1:} Based on their net income in the last quarter, should I invest in Pepsi or Coca Cola?\\
\newline \textbf{Financial Agent Formatted Data Request:}
%\newline Companies: PepsiCo, Inc.; Coca-Cola Co\\ Metrics: Net Income; Total Revenue; Cost of Revenue; Operating Expense; Depreciation and Amortization; Interest Expense\\Dates: last quarter\\
%\newline \textbf{Formatted Query as a data request for Data Module:} 
{\fontfamily{qcr}\selectfont
 (PepsiCo, Inc.; Coca-Cola Co) (Net Income; Total Revenue; Cost of Revenue; Operating Expense; Depreciation and Amortization; Interest Expense) \\(31/3/2023 - 30/6/2023)\\
}\\
\textbf{Data Module Response:} Retrieved tabular and news data 
\vspace{-2mm}
\begin{center}
 \begin{tabular}{|p{1.1cm}|p{1.8cm}|p{1.8cm}|p{1.8cm}|p{1.8cm}|p{2.0cm}|p{1.8cm}|}
 \hline
 \multicolumn{7}{|c|}{\scriptsize Table: 31/3/2023 - 30/6/2023}\\
 \hline
 \textbf{\tiny Entity}&
 \textbf{\tiny Net Income\newline (in \newline thousands) \newline (30/3/2023-29/6/2023) \newline (Quarterly)} &
 \textbf{\tiny Total \newline Revenue \newline (in \newline thousands) \newline (30/3/2023-29/6/2023) \newline (Quarterly)} &
 \textbf{\tiny Cost \newline of Revenue \newline (in \newline thousands)\newline (30/3/2023-29/6/2023) \newline (Quarterly)} &
 \textbf{\tiny Operating Expense\newline (in \newline thousands) \newline (30/3/2023-29/6/2023) \newline (Quarterly)} &
 \textbf{\tiny Depreciation and \newline Amortization \newline (in thousands)\newline (30/3/2023-29/6/2023) \newline (Quarterly)} &
 \textbf{\tiny Interest Expense \newline (in \newline thousands)\newline (30/3/2023-29/6/2023) \newline (Quarterly)}\\
 \hline
 \scriptsize PepsiCo, Inc. &\scriptsize [1932000, 2748000] &\scriptsize [17846000, 22322000]&\scriptsize [7988000, 10121000]&\scriptsize [7229000, 8542000]	&\scriptsize [706000, 810000]&\scriptsize [200000, 201000]\\ 
 \hline
 \scriptsize Coca-Cola Co &\scriptsize [3107000, 2547000]&\scriptsize [10980000, 11972000] &\scriptsize [4317000, 4912000]&\scriptsize [3296000, 4659000]&\scriptsize [286000, 281000] &\scriptsize [372000, 374000]\\ 
 \hline
 % \small Coca-Cola Co & \small [9771000,9542000]\\
 % \hline
\end{tabular}

\end{center}
[News: While PepsiCo's current performance is strong,
 some analysts expressed concern about the potential impact of inflation on consumer spending habits in the coming $\ldots$]\\
\textbf{\\Enriched query provided to ChatGPT}\\
\\You are an expert financial advisor. You will be provided with financial data and a financial query, and you have to answer the query based on the analysis of the data.\\
\\Financial Data:\\
Net Income (in thousands) (30/3/2023 - 29/6/2023) (Quarterly):\\
PepsiCo, Inc. : 1932000, 2748000\\
Coca-Cola Co : 3107000, 2547000\\
$\ldots$\\
Interest Expense (in thousands) (30/3/2023 - 29/6/2023) (Quarterly):
\\PepsiCo, Inc. : 200000, 201000
\\Coca-Cola Co : 372000, 374000
\\News:  While PepsiCo's current performance is strong,
 some analysts expressed concern about the potential impact of inflation on consumer spending habits in the coming $\ldots$\\
\\Financial Query:
Based on their net income in the last quarter, should I invest in Pepsi or Coca Cola?
\\ \\
\textbf{ChatGPT response:}\\
\begin{mdframed}[style=outbox]
To determine whether you should invest in PepsiCo or Coca-Cola based on their net income in the last quarter, let's analyze the financial data:\\
\\Net Income (in thousands) for the last quarter (30/3/2023 - 29/6/2023):
\\- PepsiCo, Inc.: 2,748,000
\\- Coca-Cola Co: 2,547,000\\
\\Now, let's calculate some key financial ratios and provide additional insights:\\
\\1. **Profit Margin**: This ratio measures the percentage of revenue that turns into profit.
   \\- PepsiCo, Inc. Profit Margin = (Net Income / Total Revenue) * 100
   \\- PepsiCo, Inc. Profit Margin = (2,748,000 / 22,322,000) * 100 = 12.31\%
   \\- Coca-Cola Co Profit Margin = (Net Income / Total Revenue) * 100
   \\- Coca-Cola Co Profit Margin = (2,547,000 / 11,972,000) * 100 = 21.27\%\\
   \\2. **Comparison**: Coca-Cola Co has a higher profit margin (21.27\%) compared to PepsiCo, Inc (12.31\%) for the last quarter. This indicates that Coca-Cola is more efficient at converting its revenue into profit during this period.\\
   \\3. **Considerations**: While profit margin is an important metric, it's essential to consider other factors like the company's growth prospects, competitive positioning, dividend yield, and long-term financial stability when making investment decisions. You should also consider your investment goals, risk tolerance, and portfolio diversification strategy.\\
\\4. **Conclusion**: Based solely on the net income in the last quarter, Coca-Cola Co appears to have a stronger performance in terms of profit margin. However, this is just one aspect of the investment decision. It's recommended to conduct a more comprehensive analysis and consider the broader financial health and strategic outlook of both companies before making an investment decision. Remember that investing involves risks, and it's advisable to consult with a financial advisor and conduct thorough research before making any investment decisions. Additionally, past performance may not guarantee future results, so it's essential to consider a company's overall financial health and long-term prospects.
\end{mdframed}
\newpage
\textcolor{black}{Next we look at examples, where knowledge-grounding is not done and ChatGPT is used directly without the Financial Agent. The queries were posed on June 2025, with ChatGPT having access to the web. It tries to retrieve the best available information to answer the query. However, in the context of Finance, the best available online information may not be current or relevant. In the absence of reliable information, the platform will be of limited value to experts.}
%\textcolor{black}{Query: What is the current stock price of Apple?}
%\textcolor{black}{ChatGPT:} 
%\begin{center}
%    \includegraphics[width=15cm]{stockprice.png}
%\end{center}
%\textcolor{black}{As of the latest available data, Apple Inc. (NASDAQ: AAPL) is trading at 209.33 USD, reflecting a slight decrease of 0.91\% from the previous close. Please note that stock prices are subject to change during market hours. For the most current information, consider checking financial news platforms or stock market apps.}
%\textcolor{black}{Comment: ???}
\\ \\
\noindent\textbf{\textcolor{black}{Example Query 2:} }\textcolor{black}{Please provide quarterly net sales for Reliance Industries Ltd for the last 12 quarters.}
\\ \\
\textcolor{black}{\noindent\textbf{ChatGPT with web access:} }
\begin{mdframed}[style=outbox]
Here is a summary of the quarterly revenue for Reliance Industries Limited (RIL) over the last 12 quarters:

\begin{center}
 \begin{tabular}{p{6cm} p{6cm} }
 \textbf{\scriptsize Quarter Ending} & \textbf{\scriptsize Revenue (\rupee~Crores)}\\
 \hline
 \small Mar 2025 & \small \rupee2,61,388\\
 \small Dec 2024 & \small \rupee2,43,865\\
 \small Sep 2024 & \small \rupee2,35,481\\
 \small Jun 2024 & \small \rupee2,36,217\\
 \small Mar 2024 & \small \rupee2,40,715\\
 \small Dec 2023 & \small \rupee2,27,970\\
 \small Sep 2023 & \small \rupee2,34,956\\
 \small Jun 2023 & \small \rupee2,10,831\\
 \small Mar 2023 & \small \rupee2,16,265\\
 \small Dec 2022 & \small \rupee2,28,955\\
 \small Sep 2022 & \small \rupee2,40,000\\
 \small Jun 2022 & \small \rupee2,30,000\\
\end{tabular}
\end{center}
\textit{Note: The revenue figures for Dec 2022, Sep 2022, and Jun 2022 are approximate, as specific data for these quarters was not available in the provided sources.}\\\\
The data indicates a general upward trend in RIL's revenue over the past three years, with notable growth in the most recent quarters. For instance, the revenue increased from \rupee2,16,265 crore in Mar 2023 to \rupee2,61,388 crore in Mar 2025, reflecting a year-on-year growth.\\\\
If you require more detailed financial metrics such as EBITDA, net profit, or segment-wise performance, feel free to ask!
\end{mdframed}

\noindent\textbf{\textcolor{black}{Comment:} }\textcolor{black}{Apart from March 2025, most of the numbers are wrong. Additionally, the model claims, it has provided approximate numbers for June, September and December 2022 as it was unable to find information online.}
\newpage
\noindent\textbf{\textcolor{black}{Example Query 3:} }\textcolor{black}{Please provide quarterly revenue figures for General Motors for the last 12 quarters.}
\\\\
\textcolor{black}{\noindent\textbf{ChatGPT with web access:} }
\begin{mdframed}[style=outbox]
Here are General Motors' quarterly \textbf{total revenue} figures (automotive + GM Financial) for the last 12 reported quarters:

\begin{center}
 \begin{tabular}{p{5cm} p{7cm} }
 \textbf{\scriptsize Quarter } & \textbf{\scriptsize Revenue (USD billions)}\\
 \hline
 \small Q1 2025 (ended Mar 31)& \small **44.0**\\
 \small Q4 2024 (ended Dec 31)& \small **47.7**\\
 \small Q3 2024 (ended Sep 30)& \small $\sim$**43.0**\\
 \small Q2 2024 (ended Jun 30)& \small (not in press release; implies $\sim$44)\\
 \small Q1 2024 (ended Mar 31)& \small (not explicitly published; similar to $\sim$43)\\
 \small Q4 2023 (ended Dec 31)& \small $\sim$**43.0**\\
 \small Q3 2023 (ended Sep 30)& \small (likely around **40-41**)\\
 \small Q2 2023 (ended Jun 30)& \small (approximately **38-39)\\
 \small Q1 2023 (ended Mar 31)& \small (approx. **35-36**)\\
 \small Q4 2022 (ended Dec 31)& \small (approx. **37-38**)\\
 \small Q3 2022 (ended Sep 30)& \small (approx. **37**)\\
 \small Q2 2022 (ended Jun 30)& \small (approx. **36**)\\
\end{tabular}
\end{center}
To get precise numbers for Q2 2024-Q2 2022, more detailed quarterly reports are needed. The figures above are estimated based on publicly available sources. If you'd like, I can look up each of those specific quarters via GM's SEC filings. 
\end{mdframed}
\begin{mdframed}[style=inbox]
Please look up in the SEC filings and provide accurate numbers.
\end{mdframed}
\begin{mdframed}[style=outbox]
I wasn't able to find a single table listing all past 12 quarterly revenues directly from SEC filings online. To provide accurate numbers, we'll need to manually pull them from each quarterly 10-Q report. I can guide you through a step-by-step approach to retrieve and compile this data, unless you prefer an alternative:\\
\\
\textbf{Option 1: I walk you through how to download each quarterly 10-Q from GM's Investor Relations website and extract the "Total revenue" or "Net sales and revenue" figure for each quarter.}\\
\\
\textbf{Option 2: I summarize what I can find publicly online (e.g. in GM's press releases or in quarterly earnings summaries) and flag where exact figures from 10-Qs are still missing.}\\
\\
Let me know which approach you'd prefer---or if you'd like me to proceed directly and list out each quarter by accessing the PDFs one-by-one.
\end{mdframed}

\noindent\textbf{\textcolor{black}{Comment:} }\textcolor{black}{It is clear that the model is unable to handle this query because of the absence of a module which can provide relevant data. Further interactions with ChatGPT are also not helpful in this context. Our proposed approach exploits it with the help of Financial Agent.}
\\\\
\textcolor{black}{Next, we compare the performance of ChatGPT with web access enabled and ChatGPT with data access through Financial Agent and Data Module. We use the same query to demonstrate the performance.}
\\\\
\noindent\textcolor{black}{\textbf{Example Query 4: } Based on Magnus Retail Limited's EBITDA for last 4 quarters, is it a suitable investment?}
\\\\
\textcolor{black}{\noindent\textbf{ChatGPT with web access:} }
\begin{mdframed}[style=outbox]
To assess whether \textbf{Magnus Retail Limited} is a suitable investment based on its \textbf{EBITDA (Earnings Before Interest, Taxes, Depreciation, and Amortization)} over the \textbf{last 4 quarters}, I need the actual \textbf{quarter-wise EBITDA figures} or a chart/image containing them.\\\\
Could you please provide:\\
• The \textbf{EBITDA values for each of the last 4 quarters}, or\\
• A screenshot or table containing that information?\\\\
Once I have that, I'll:\\
1. Analyze the trend (growth/decline/stability).\\
2. Check EBITDA margins if revenue is available.\\
3. Offer a conclusion on investment suitability based on operating performance.\\\\
Let me know if you also want to consider other financials like net income, debt, or valuation ratios.    
\end{mdframed}
\noindent\textcolor{black}{\textbf{Enriched Query provided to ChatGPT: } \\\\You are an expert financial advisor. You will be provided with financial data and a financial query, and you have to answer the query based on the analysis of the data.
\\
Financial Data: EBITDA (in thousands) (INR) (Quarterly):
\\
Magnus Retail Limited:
\\
31/3/2025: 3092
\\
31/12/2024: -1003
\\
30/9/2024: -195
\\
30/6/2024: -569
\\
Financial News: None
\\
Financial Query: Based on Magnus Retail Limited’s EBITDA for last 4 quarters, is it a suitable investment?}
\newpage
\textcolor{black}{\noindent\textbf{ChatGPT Response to Enriched Query:} }
\begin{mdframed}[style=outbox]
 \small To assess whether \textbf{Magnus Retail Limited} is a suitable investment based solely on its \textbf{EBITDA (Earnings Before Interest, Taxes, Depreciation, and Amortization)} for the last four quarters, we need to examine the trend and evaluate its financial implications.\\\\
\textbf{\large EBITDA Trend (in INR '000):}
\begin{center}
 \begin{tabular}{p{4cm} p{4cm} p{4cm} }
 \textbf{\scriptsize Quarter Ending} & \textbf{\scriptsize EBITDA} & \textbf{\scriptsize Change (QoQ)}\\
 \hline
 \small 31/3/2025 & \small 3.092 & \small +4,095 \\
 \small 31/12/2024 & \small -1,003 & \small -808 \\
 \small 30/9/2024 & \small -195 & \small +374 \\
 \small 30/6/2024 & \small -569 & \small --- \\
 
\end{tabular}
\end{center}
\textbf{\large Key Observations:}\\\\
\small\textbf{1. Sustained Negative EBITDA Prior to Q4 FY25:}\\
\small The company reported negative EBITDA for three consecutive quarters, reflecting operational challenges and possibly inefficient cost structures or declining revenues.\\
\small \textbf{2. Strong Recovery in Q4 FY25:}\\
\small The EBITDA rose sharply to 3,092 thousand INR from -1,003 thousand INR the previous quarter. This turnaround may signal improvements such as increased sales, better cost management, or other operational efficiencies. However, without financial disclosures, it's difficult to assess whether this change is structural or temporary.\\
\small \textbf{3. Volatile Performance:}\\
\small The pattern suggests instability in core operations. A single quarter of positive EBITDA following a period of losses should be interpreted cautiously, especially in the absence of accompanying financial news or context.\\\\
\textbf{\large Assessment of Investment Suitability (Based on EBITDA):}\\\\
\small \textbf{Positives:}\\
\small • Evidence of operational improvement in the most recent quarter.\\
\small • Potential early signs of turnaround.\\
\small \textbf{Negatives:}\\
\small • Prior negative EBITDA indicates weak recent performance.\\
\small • Lack of transparency or supporting financial details makes it difficult to determine if the recent improvement is sustainable.\\
\small • High volatility in EBITDA suggests ongoing business risks.\\\\
\textbf{\large Conclusion:}\\\\
\small \textbf{Magnus Retail Limited is not yet a suitable investment based solely on EBITDA.}\\
\small • Multiple consecutive quarters of positive EBITDA\\
\small • More detailed financial disclosures (e.g., revenue, net income, cash flow)\\
\small • Clarification on what drove the Q4 improvement\\\\
\small \textbf{Recommendation:} Add Magnus Retail Limited to a watchlist and monitor its performance in upcoming quarters. Avoid investing based solely on a single quarter of positive results following a series of losses.
\end{mdframed}
\end{document}